\documentclass[10pt]{article}
\usepackage{graphicx}
\usepackage{amssymb}
\usepackage{amsmath, setspace, amssymb}
\usepackage{pifont}
\usepackage{float}
\usepackage[labelfont=bf]{caption}
\usepackage{color}
\usepackage{multicol}
\usepackage[numbers,sort&compress,super,comma]{natbib} 
\usepackage{filecontents} 

\newcommand{\ba}[1]{ \begin{align}#1\end{align} }

\newcommand{\BAMAPbILong}{(C$_4$H$_9$NH$_3$)$_2$(CH$_3$NH$_3$)$_{n-1}$Pb$_n$I$_{3n+1}$}
\newcommand{\BAPbI}{BA$_2$PbI$_4$}
\newcommand{\BAPbILong}{(C$_4$H$_9$NH$_3$)$_2$PbI$_4$}

\newcommand{\ket}[1]{|#1\rangle}
\newcommand{\bra}[1]{\langle#1|}

\newcommand{\kpar}{\mathbf{k}_{||}}
\newcommand{\kp}{k_{||}}

\newcommand*{\citen}[1]{%
  \begingroup
    \romannumeral-`\x 
    \setcitestyle{numbers}%
    \cite{#1}%
  \endgroup   
}

\begin{document}
\newcommand{\printtitle}{\textbf{Bright magnetic dipole radiation from two-dimensional lead-halide perovskites}}
\newcommand{\printauthors}{
Ryan A. DeCrescent,\textsuperscript{1} 
Naveen R. Venkatesan,\textsuperscript{2} 
Clayton J. Dahlman,\textsuperscript{2}  
Rhys M. Kennard,\textsuperscript{2}  
Xie Zhang,\textsuperscript{2}  
Wenhao Li,\textsuperscript{3},
Xinhong Du,\textsuperscript{4}
Michael L. Chabinyc,\textsuperscript{2}  
Rashid Zia\textsuperscript{3}  
and
Jon A. Schuller\textsuperscript{*,4} 
}
\newcommand{\printaffiliations}{
\textsuperscript{1}\emph{Department of Physics, University of California Santa Barbara, Santa Barbara, CA 93106, USA}
\\
\textsuperscript{2}\emph{Materials Department, University of California Santa Barbara, Santa Barbara, CA 93106, USA}
\\
\textsuperscript{3}\emph{Brown University, School of Engineering, Providence, Rhode Island 02912, USA.}
\\
\textsuperscript{4}\emph{Department of Electrical and Computer Engineering, University of California, Santa Barbara, California 93106, United States}
}
\newcommand{\printcorrespondence}{*Correspondence should be addressed to J.A.S. (email: jonschuller@ece.ucsb.edu)}

\begin{center}
\printtitle \\[10pt]
\printauthors \\[10pt]
\printaffiliations \\[10pt]
\printcorrespondence
\end{center}

\section*{ABSTRACT}
Light-matter interactions in semiconductor systems are uniformly treated within the electric dipole (ED) approximation, as multipolar interactions are considered ``forbidden". Here, we demonstrate that this approximation inadequately describes light emission in novel two-dimensional hybrid organic-inorganic perovskite materials (2D HOIPs) --- a class of solution processable layered semiconductor with promising optoelectronic properties. Consequently, photoluminescence (PL) spectra become strongly dependent on the experimental geometry, a fact that is often overlooked, though critical for correct optical characterization of materials. Using energy-momentum and time-resolved spectroscopies, we experimentally demonstrate that low-energy sideband emission in 2D HOIPs exhibits a highly unusual, multipolar polarization and angle dependence. Using combined electromagnetic and quantum-mechanical analyses, we attribute this radiation pattern to an out-of-plane oriented magnetic dipole transition arising from the 2D character of the excited and ground state orbitals. Symmetry arguments point toward the presence of significant inversion symmetry-breaking mechanisms that are currently under great debate. These results provide a new perspective on the origins of unexpected sideband emission in HOIPs, clarify discrepancies in previous literature, and generally challenge the paradigm of ED-dominated light-matter interactions in novel optoelectronic materials.

\newpage
\begin{multicols}{2}
\section*{BODY}
\begin{figure}[H]
	\centering
      \includegraphics[width=0.42\textwidth]{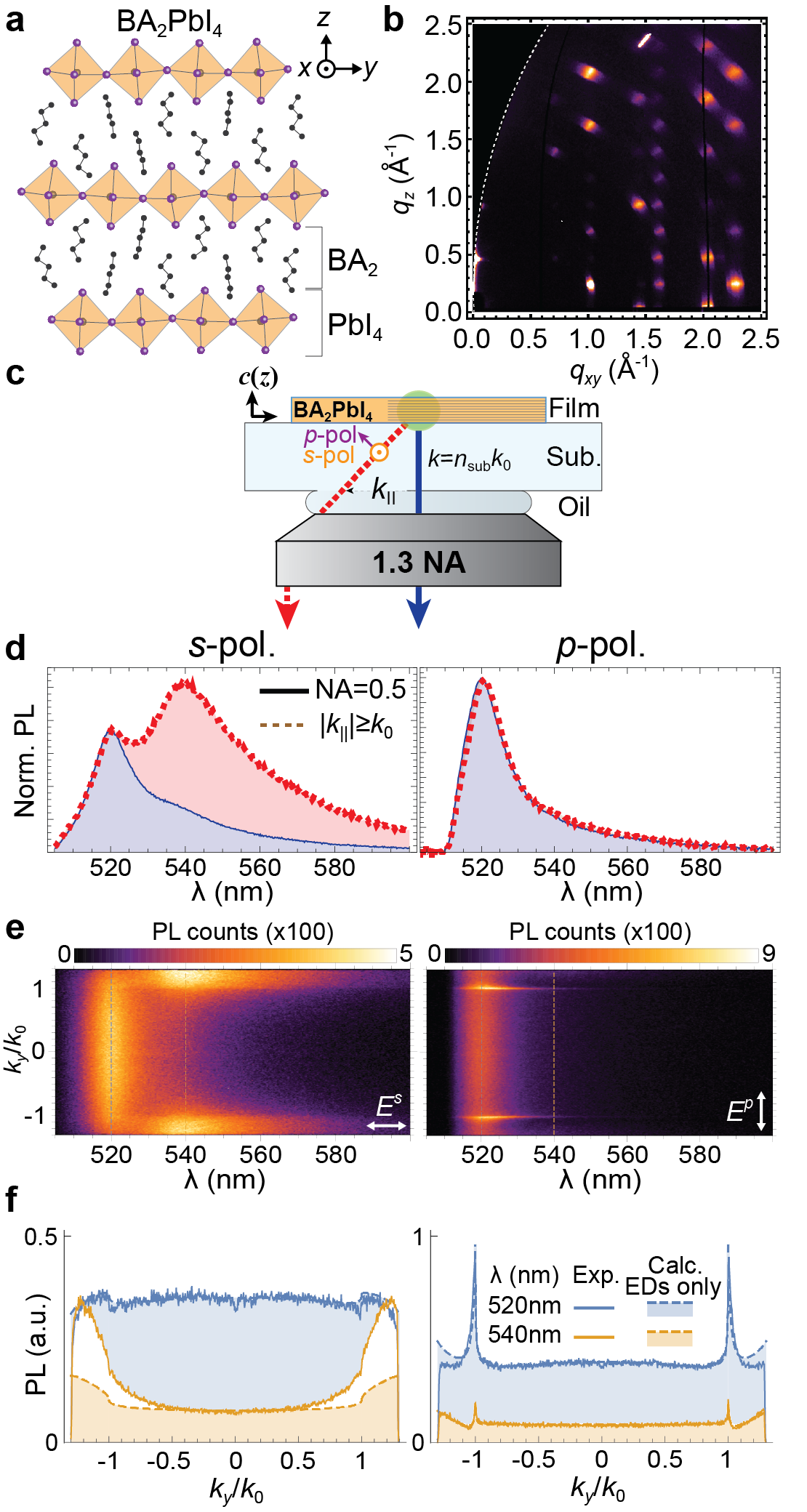}
	\caption{
	\textbf{\emph{Structure and energy-momentum photoluminescence spectra of \BAPbI.}} 
	\footnotesize
	(\textbf{a}) Schematic crystal structure of \BAPbI; sheets of semiconducting metal-halide (Pb-I) layers are separated by long alkylammonium (BA) molecules which act as an electrically insulating layer. Molecular and octahedral orientations as presented here are intended only to be qualitative.
	(\textbf{b}) Experimental GIWAXS patterns of a spin-cast \BAPbI \ thin film. Thin films exhibit a vertically layered morphology with repeated PbI$_4$ monolayers separated in the $z$-direction by BA$_2$ spacer layers. 
	(\textbf{c}) Experimental geometry: Momentum- ($k_{||}$) and polarization-dependent PL spectra are collected from within the substrate by an oil-immersion 1.3 NA objective. 
	(\textbf{d}) $s$-Polarized (left) and $p$-polarized (right) photoluminescence spectra of a \BAPbI \ thin film (61nm) \ as collected at two very different regions in momentum space: $|k_{||}|$$<$0.5$k_0$ (solid blue) and $|k_{||}|$$>$$k_0$ (dashed red). A secondary, very strong peak at 540 nm is observable in only $s$-polarized spectra at large $k_{||}$, indicating a highly oriented multipolar transition. PL traces are normalized to be equivalent at 520 nm. 
	(\textbf{e}) $s$-Polarized (left) and $p$-polarized (right) energy-momentum spectra from which the photoluminescence spectra of Fig. \ref{fig:mPL}d were taken. The multipolar emission is readily observable in $s$-polarized spectra as two bright lobes at $\lambda$=540 nm in regions with $|k_{||}|$$>$$k_0$. 
	(\textbf{f}) Momentum-space line cuts at 520 nm (blue) and 540 nm (orange) for $s$-polarization (left) and $p$-polarization (right). Theoretical traces (dashed lines) assuming only oriented electric dipoles agree poorly with 540 nm $s$-polarized line cuts, showing that the bright high-$k_{||}$ emission cannot originate from an electric dipole transition.}\label{fig:mPL}
\end{figure}

Hybrid organic-inorganic perovskites (HOIPs) are a rapidly burgeoning class of semiconductors, offering benefits of solution processability, outstanding optoelectronic properties, and the ability to form both three-dimensional (3D) and quantum confined structures.\cite{brenner_hybrid_2016,herz_charge-carrier_2016,zhang_metal_2016,garcia_de_arquer_solution-processed_2017} 
Of particular interest for photonic applications, 2D HOIPs incorporating large alkylammonium molecules (e.g., butylammonium lead iodide, \BAPbI; Fig. \ref{fig:mPL}a) offer additional structural and quantum degrees of freedom, providing continuously tunable band-gap energies and desirable narrow excitonic luminescence.\cite{stoumpos_ruddlesdenpopper_2016,tsai_high-efficiency_2016,pedesseau_advances_2016,kamminga_confinement_2016,congreve_tunable_2017,quan_perovskites_2018} 
Indeed, researchers have demonstrated stable light-emitting diodes\cite{congreve_tunable_2017,tsai_stable_2018,jia_super_2018} (LEDs) with high quantum efficiencies,\cite{tsai_stable_2018,yang_efficient_2018} as well as low-threshold optical gain.\cite{raghavan_low-threshold_2018}  However, numerous studies reveal anomalous and undesirable absorption and emission sideband features in both 2D\cite{mitzi_synthesis_1996, gauthron_optical_2010,dohner_intrinsic_2014,cao_2d_2015,straus_direct_2016,blancon_scaling_2018,tsai_stable_2018, gong_electronphonon_2018,walters_quantum-confined_2018} and 3D HOIPs,\cite{kong_characterization_2015,dar_origin_2016,wang_indirect_2017,wu_indirect_2019} leading to ``intrinsic white-light emission" that is detrimental to color purity. 

Insofar as 2D HOIPs are comprised of alternating semiconducting (metal halide) and insulating (organic) layers, they are recognized as ``natural" quantum-well semiconductor structures.\cite{ishihara_exciton_1989,mitzi_conducting_1995,tanaka_two-dimensional_2002,tanaka_bandgap_2003,even_understanding_2014,yaffe_excitons_2015,stoumpos_ruddlesdenpopper_2016,kamminga_confinement_2016,sapori_quantum_2016,blancon_extremely_2017} Accordingly, light-matter interactions in HOIPs are treated by analogy to widely studied conventional semiconductors, such as GaAs.\cite{ishihara_exciton_1989,tanaka_electronic_2005,even_electronic_2012,even_importance_2013,even_analysis_2014,yu_effective-mass_2016,fu_neutral_2017,becker_bright_2018} Namely, optical transitions are assumed to be completely described by the lowest-order term in a multipolar expansion of the Hamiltonian --- the electric dipole (ED) interaction --- as higher-order interactions are conventionally slow enough to be considered ``forbidden".\cite{cowan_theory_1981,bastard_wave_1988,bunker_molecular_2006} Operating within this conventional framework, researchers have attempted to identify the origin of anomalous low-energy sideband features in 2D and 3D HOIPs, arriving at interpretations ranging from bound- or bi-exciton emission\cite{ishihara_optical_1990,fujisawa_excitons_2004,chong_dominant_2016,booker_vertical_2018} to strong phonon-carrier interactions,\cite{hong_dielectric_1992,gauthron_optical_2010,dohner_intrinsic_2014,straus_direct_2016,iaru_strong_2017,gong_electronphonon_2018,straus_direct_2016} and in other cases remaining utterly unexplained.\cite{mitzi_synthesis_1996} Recent evidence of more ``exotic" and fundamentally interesting physics, such as strong Rashba and Dresselhaus couplings\cite{zhai_giant_2017,kepenekian_rashba_2017} and an unconventional exciton fine structure,\cite{fu_neutral_2017,becker_bright_2018,chen_composition-dependent_2018} reflect the interplay of strong spin-orbit coupling, structural complexity,\cite{zhang_three-dimensional_2018,zhang_first-principles_2018} and the possibility of significant dynamic symmetry-breaking mechanisms.\cite{niesner_structural_2018,frohna_inversion_2018,zheng_structural_2018,wu_indirect_2019} Regardless of the mechanism, the observed strength of this sideband emission varies dramatically and curiously between these reports, tempting a judgement of sample quality.

Here, using energy-momentum spectroscopy, we demonstrate that the low-energy sideband emission in 2D HOIPs exhibits a peculiar polarization and angle dependence which is intrinsically related to the 2D nature of the material. Energy-momentum spectroscopies have surfaced as powerful techniques for identifying anisotropic and multipolar optical phenomena in thin film systems.\cite{taminiau_quantifying_2012,schuller_orientation_2013,karaveli_time-resolved_2013,li_quantifying_2014,brown_morphology-dependent_2016,decrescent_model-blind_2016,brown_enhancing_2017,scott_directed_2017,fieramosca_tunable_2018} For example, by analyzing momentum- and polarization-dependent luminescence profiles, researchers have quantified distinct magnetic dipole (MD) and ED optical transitions in lanthanide\cite{taminiau_quantifying_2012,li_quantifying_2014} and transition-metal ions,\cite{karaveli_time-resolved_2013} and have identified distinct interlayer excitons in organic semiconductor thin films.\cite{schuller_orientation_2013} Using combined electromagnetic and quantum-mechanical analyses, we attribute the anomalous PL from 2D HOIPs to an out-of-plane MD transition from an excitonic state. Symmetry considerations point toward the existence of electronic symmetries that are not revealed in static DFT calculations, supporting the presence of significant dynamical symmetry breaking mechanisms that have recently been suggested.\cite{niesner_structural_2018,frohna_inversion_2018,zheng_structural_2018,wu_indirect_2019} This demonstrates the critical role of the experimental geometry in the optical characterization of novel 2D materials, and likely accounts for the wide variation of previous interpretations of sideband emission in HOIPs.

While the features described herein seem to be generic to the material system, including 3D HOIPs, we exploit the highly oriented and 2D nature of \BAPbI \ (Fig. \ref{fig:mPL}a) to relate the unusual radiation patterns to the underlying quantum mechanical origins. Figure \ref{fig:mPL}b shows measured grazing incidence wide-angle X-ray scattering (GIWAXS) patterns from a spin-cast thin film of \BAPbI. This data is consistent with a vertically layered structure in which PbI$_4$ monolayers are separated by BA$_2$ spacer layers in the out-of-plane ($z$) direction.\cite{venkatesan_charge-carrier_2018,venkatesan_phase_2018} The wavelength-dependent uniaxial complex refractive index (Supplementary Fig. 1) of these samples was determined by a combination of momentum-resolved reflectometry\cite{decrescent_model-blind_2016} and variable-angle spectroscopic ellipsometry (Methods; Supplementary Information S1). Films exhibit a predominant in-plane (IP) optical response with weak out-of-plane (OP) dispersion, similar to high-quality single crystals\cite{ishihara_optical_1990,guo_hyperbolic_2018} and consistent with the vertically layered orientation as determined from GIWAXS. By the nature of the spin-casting method, the resulting films are polycrystalline and thus rotationally isotropic over microscopic length scales relevant for the measurements performed here ($\sim$100 $\mu$m).\cite{venkatesan_charge-carrier_2018} The highly oriented and rotationally invariant thin-film structure facilitates detailed analysis of momentum- and polarization-dependent optical phenomena using the experimental geometry illustrated in Fig. \ref{fig:mPL}c. However, as we will describe later, the same features are observed from high-quality single crystals and our conclusions are thus not restricted to spin-cast films. In energy-momentum spectroscopy, both the in-plane electromagnetic momentum ($\kpar$=$\langle$$k_x$,$k_y$$\rangle$) and wavelength ($\lambda$) distribution of polarized reflection or luminescence are simultaneously measured by imaging the Fourier plane (i.e., back focal plane; BFP) of a collection objective onto the entrance slit of an imaging spectrometer. Polarized spectra are acquired by placing a linear polarizer in the collection path oriented either perpendicular ($s$) or parallel ($p$) to the entrance slit ($y$-axis). Figures \ref{fig:mPL}d-f show $s$- (left) and $p$-polarized (right) energy- and momentum-resolved PL measured from a 61 nm spin-cast film of \BAPbI \ at room temperature (Methods). 

As seen in Figure \ref{fig:mPL}d, $s$-polarized PL spectra collected above the total-internal-reflection condition $|k_{||}|$=$|k_y|$$>$$k_0$ (red dashed) are dramatically different than conventional spectra as collected by a 0.5 NA objective ($|k_{||}|\leq0.5k_0$; blue). Both $s$- (left) and $p$-polarized (right) spectra exhibit the well-established excitonic emission feature at 520 nm. However, a distinct emission feature at 540 nm ($\sim$90 meV energy difference), apparent as a subtle shoulder at low momenta (blue), 

\noindent becomes the dominant emission feature at high momenta (red dashed). In contrast, this emission feature is virtually absent at \emph{all} momenta in $p$-polarized spectra. This difference in momentum distributions is even more striking when we examine the full energy-momentum spectra (Figure \ref{fig:mPL}e; polarization orientation indicated by white arrows).  The low-energy emission feature is readily observed as two bright lobes in the $s$-polarized spectra at $|k_{||}|$$>$$k_0$. Importantly, the $s$-polarized emission features at 520 nm and 540 nm have markedly distinct curvature in $k$-space (Fig. \ref{fig:mPL}f; left). The emission feature at 520 nm (blue) is maximal near normal ($k_{||}$=0) and exhibits a slow roll-off into higher momenta; the opposite is true for emission at 540 nm (orange). This illustrates the critical role of the experimental geometry on the inferred significance of this emission (Supplementary Fig. 2); because this light is emitted at highly oblique angles with respect to the crystal axis, it would be ordinarily be trapped (i.e., wave-guided) within the high-index substrate, or within the material itself, and thus only weakly collected with conventional PL geometries (e.g., blue lines in Fig. \ref{fig:mPL}d). This is a general concern for 2D materials, since the polarization and directionality of the emitted light is interacting with highly anisotropic electronic wave functions. This may also have significant implications for optoelectronic devices, especially LEDs, based upon 2D HOIPs; the bandwidth of emission can be enhanced by strategically collecting and redirecting this highly oblique light. Alternatively, the narrow excitonic emission can be retained by allowing this oblique low-energy light to remain trapped in wave-guided modes.

Calculated $s$-polarized PL counts (Supplementary Information S2), assuming only the existence of ED emission, are shown in Fig. \ref{fig:mPL}f (dashed lines) at both 520 nm (blue) and 540 nm (orange). Experimental PL at the 520 nm (primary exciton) emission is accurately described by the ED theory, consistent with an ED-allowed excitonic transition involving $\Gamma_6^-$ CBM and $\Gamma_6^+$ VBM states.\cite{umebayashi_electronic_2003,even_electronic_2012} The 540 nm emission feature, however, is poorly modeled by the same theory. In contrast, $p$-polarized PL at both 520 nm and 540 nm are well-described by ED-only theory. While low-energy emission shoulders have been observed in conventional PL spectra of both inorganic and hybrid organic-inorganic perovskites,
\cite{ishihara_optical_1990,mitzi_synthesis_1996,fujisawa_excitons_2004,dohner_intrinsic_2014,straus_direct_2016,chong_dominant_2016,iaru_strong_2017,tsai_stable_2018,gong_electronphonon_2018,blancon_scaling_2018,walters_quantum-confined_2018,booker_vertical_2018,wu_indirect_2019} none of these prior works examined the polarization and momentum dependence reported here. Note that this theory completely accounts for reabsorption effects by including the complex (uniaxial) refractive index (Supplementary Fig. 1) of the material, thus excluding reabsorption as a possible explanation of the secondary peak. The strong high-$k_{||}$ emission excess observed predominantly in $s$-polarization cannot be explained by any combination of oriented ED transitions in a rotationally invariant system; a distinct ED transition centered around 540 nm would contribute with equivalent weight to both $s$- and $p$-polarized spectra, inconsistent with our data. The results presented here thus unambiguously demonstrate that this emission is multipolar in nature. 

We observe similar multipolar features at high momenta in a number of related HOIPs (Methods; Supplementary Infor-

\end{multicols}
\begin{figure}[t]
	\centering
	\includegraphics[width=0.7\linewidth]{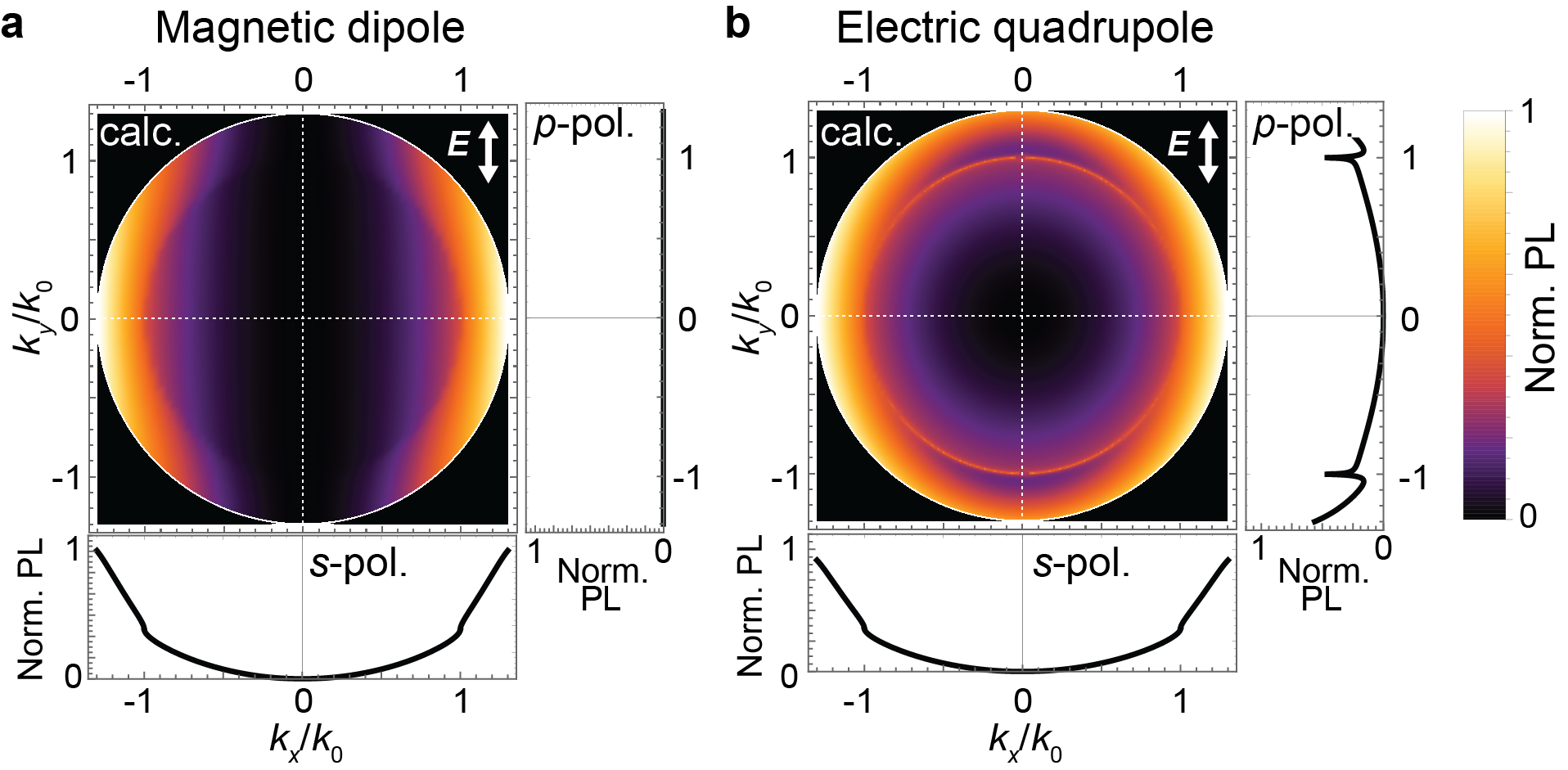}
	\caption{
	\textbf{\emph{Calculated momentum-resolved multipolar luminescence patterns at $\lambda$=540 nm.}}
	\footnotesize
	(\textbf{a}) Calculated $y$-polarized momentum-resolved luminescence patterns from an OP MD.
	(\textbf{b}) Calculated $y$-polarized momentum-resolved luminescence patterns from IP EQs. 
	$s$-Polarized and $p$-polarized traces are shown below and to the right of each 2D image, respectively.	Note that the radiation patterns from OP MDs and IP EQs differ along the $p$-polarized traces; the $s$-polarized traces are identical. All calculations were performed for a 61 nm film using optical constants from Supplementary Fig. 1.}
\label{fig:multipoles}
\end{figure}
\begin{multicols}{2}

\noindent mation S3), including drop-cast single crystals of the ``bilayer" system BA$_2$(CH$_3$NH$_3$)Pb$_2$I$_7$ (Supplementary Fig. 3), thin films of BA$_2$PbBr$_4$ (Supplementary Fig. 4), and films prepared with longer alkylammonium cations (octylammonium and dodecylammonium) (Supplementary Fig. 5). We also find that this feature is robust to sample preparation methods, and is particularly apparent in high-quality single crystals prepared with various alkylammonium cations (including phenethylammonium), and flakes mechanically exfoliated from such crystals (Supplementary Figs. 6-8). We thus conclude that this feature is general to the material system, and is robust against sample preparation methods and substitutions of both the halide and spacer molecules. 

While this feature is regularly observed at 540 nm in exfoliated single crystals, we note that, for similarly prepared \BAPbI \ thin films, this emission feature is observed with varying strength at a range of energies between $\sim$45-90 meV below the primary emission peak. Analyses of thin-film PL spectra under varying illumination and environmental conditions (Supplementary Information S4; Supplementary Fig. 9) suggest that this variation is due, in part, to the incorporation of oxygen or water\cite{aristidou_role_2015,marinova_hindered_2017} under various processing conditions. Note, however, that prolonged illumination in ambient conditions significantly \emph{reduces} the resolution of two distinct peaks, which possibly explains why this feature is so apparent in single crystals. Importantly, this indicates that this sideband does \emph{not} originate from atmospheric effects (e.g., water or oxygen), but rather seems to be hindered by them.

The highly $s$-polarized nature of the excess (i.e., non-ED) PL provides strong clues for the origins of the multipolar radiation. The candidate oriented multipoles that emit highly oblique $s$-polarized light are OP MDs and IP EQs (Supplementary Information S5; Supplementary Fig. 10). Calculated 2D $y$-polarized radiation patterns from these multipoles are presented in Fig. \ref{fig:multipoles} with $s$- and $p$-polarized $\kp$-space linecuts shown below and to the right of each 2D image, respectively. OP MDs (``MD$_z$"; Fig. \ref{fig:multipoles}a) are associated with a circulating IP electric field and thus produce only $s$-polarized radiation. However, this $s$-polarized MD$_z$ contribution is functionally identical to $s$-polarized emission from IP EQs (Fig. \ref{fig:multipoles}b; Supplementary Information S2). Although either MD or EQ terms can be used to fit the $s$-polarized emission anomaly, identifying the correct multipole term is important for determining the underlying quantum-mechanical origins. To this end, the OP MD and IP EQ can be distinguished by the fact that the OP MD emits no $p$-polarized PL; in contrast, IP EQs contribute significantly to both $s$- and $p$-polarized spectra (Supplementary Fig. 11). Critically, our data shows virtually zero PL excess in $p$-polarized spectra (Fig. \ref{fig:mPL}d), indicating that the multipolar PL is associated with a highly oriented OP MD transition. Note that, while the ED radiation patterns possess features similar to those of the EQ, the preceding discussion is focused solely upon the patterns associated with the \emph{excess} counts (i.e., those not already accounted for by the ED theory). We suspect the very subtle shoulder observed in high-$\kp$ $p$-polarized spectra is due to depolarization scattering from surface texture or slightly tilted crystallite domains in spin-cast thin films. Equivalent analyses on mechanically exfoliated single crystals (Supplementary Fig. 6), in which these imperfections are minimized, show no such $p$-polarized shoulder and confirm the MD assignment.

The energy-momentum spectra, in combination with the known material optical constants (Supplementary Fig. 1), afford the opportunity to quantify the relative intrinsic multipolar transition rates. At each wavelength, we decompose the momentum distribution of polarized PL counts, $N^{s,p}(\lambda,\kpar)$, into a linear combination of oriented EDs and OP MDs (Methods; Supplementary Information S2) according to Eqn. \ref{eqn:LDOS}:
\ba{
N^{s,p} & (\lambda,\kpar) = 
C_\text{exp} \times
\left[ 
A_{\text{ED}_\text{IP}}(\lambda) \tilde{\rho}_{\text{ED}_\text{IP}}^{s,p}(\lambda,\kpar) \ + \right. \nonumber \\
&\left. 
A_{\text{ED}_\text{OP}}(\lambda) \tilde{\rho}_{\text{ED}_\text{OP}}^{s,p}(\lambda,\kpar)
+ A_{\text{MD}_\text{OP}}(\lambda) \tilde{\rho}_{\text{MD}_\text{OP}}^{s,p}(\lambda,\kpar)
\right]
\label{eqn:LDOS}
}

\begin{figure}[H]
	\centering
	\includegraphics[width=0.9\linewidth]{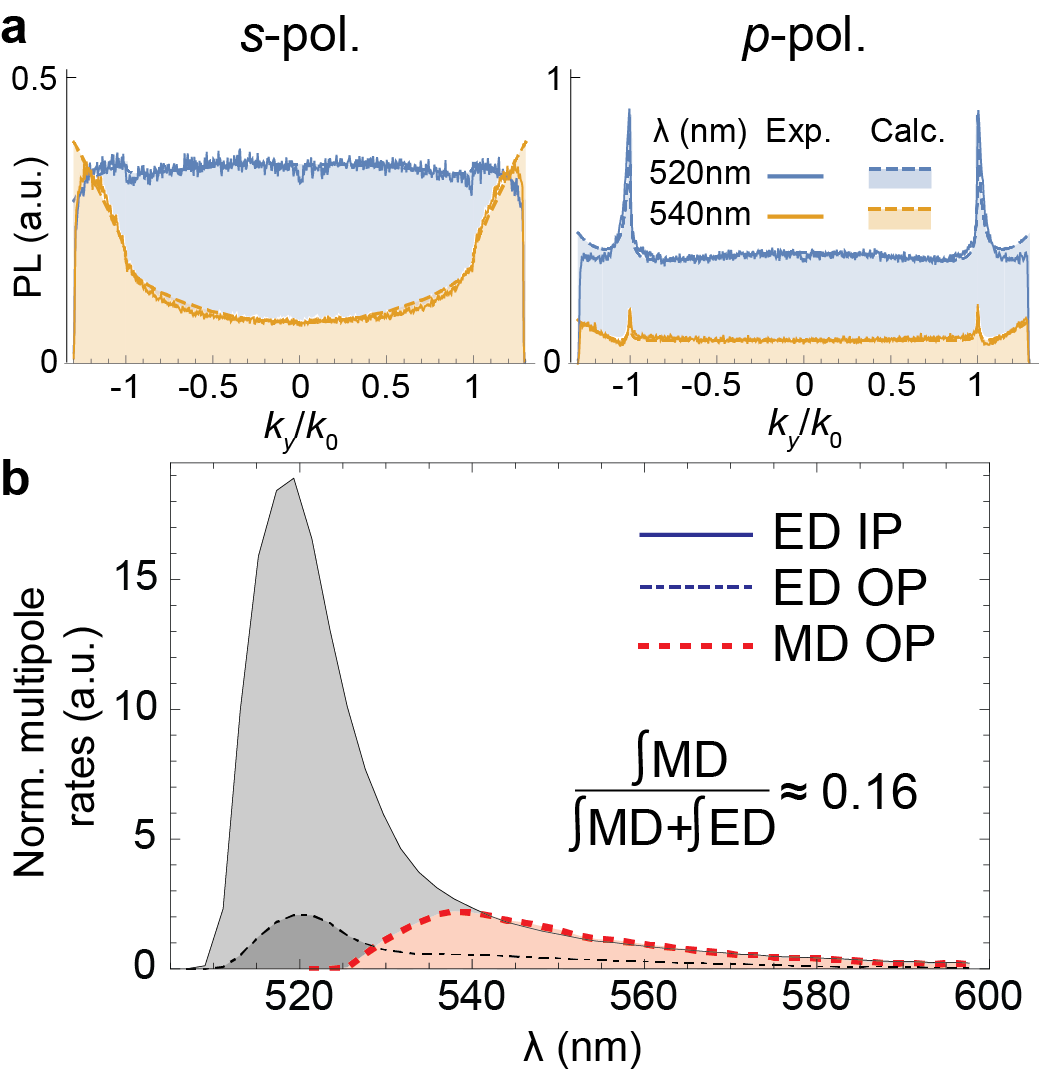}
	\caption{
	\textbf{\emph{Multipolar decomposition of energy-momentum spectra:}}  
	\footnotesize (\textbf{a}) Normalized intrinsic multipole radiation rates, $C_\text{exp}A(\lambda)$, determined from multipolar decompositions of energy-momentum spectra: IP EDs (black solid); OP EDs (black dot-dashed); OP MDs (brown dashed). (\textbf{b}) Comparison of (dashed filled) theoretical and (solid) experimental momentum-space radiation profiles at $\lambda$=520 nm (blue) and $\lambda$=540 nm (orange) including both oriented ED and MD transitions. $s$-($p$-)Polarized spectra are shown on the left (right). All traces show excellent agreement with theory once multipolar components are included in the theory.}\label{fig:decomposition}
\end{figure}

\noindent Here, the $\tilde{\rho}^{s,p}$ are the normalized local density of optical states (LDOS) into which the oriented dipoles may emit, $A(\lambda)$ are the wavelength-dependent intrinsic emission rates, and $C_\text{exp}$ is a constant factor accounting for setup-specific experimental parameters. Analytical forms for the LDOS are presented in Supplementary Information S2. Including the MD contribution, we now see that the experimental radiation patterns for both polarizations are very well represented by the theory across all wavelengths (Fig. \ref{fig:decomposition}a). The resulting decomposed spectra are presented in Fig. \ref{fig:decomposition}b. The ED and MD peaks resolved in Fig. \ref{fig:decomposition}c are observed as genuinely distinct transitions. The ED (black) and MD (dashed red) contributions exhibit similar asymmetric lineshapes characteristic of luminescence from quantum wells, suggesting that the MD transition also originates from an excitonic excited state. However, the MD emission is notably more broad with an exponential-like distribution of states below the band-gap, characteristic of the Urbach tails that are common in ionic crystals and induced by thermal disorder.\cite{dow_toward_1972,wu_indirect_2019}

Although multipolar emission is usually orders of magnitude weaker than ED emission, the multipolar contribution 

\begin{figure}[H]
	\centering
    \includegraphics[width=3.5in]{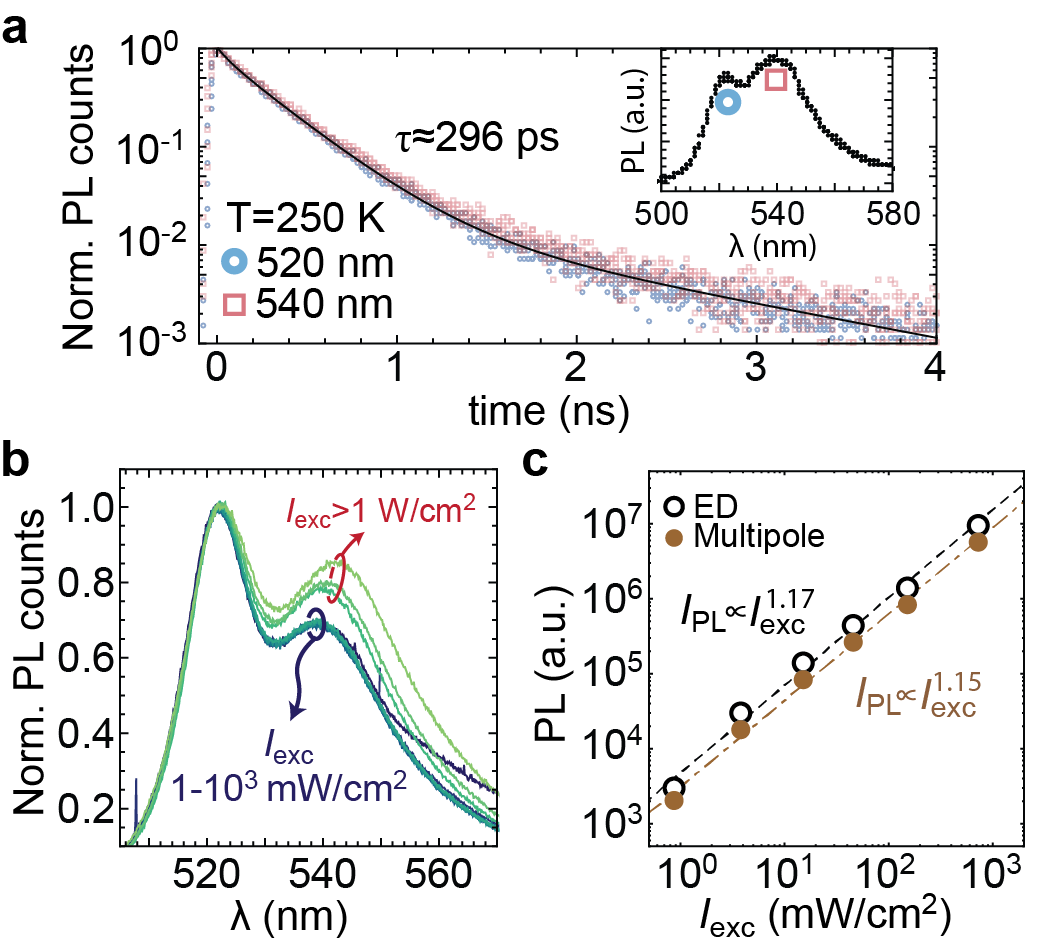}
	\caption{
	\textbf{\emph{Dynamics and pump-dependence of emission features:}}
    \footnotesize 
    (\textbf{a}) Time-resolved photoluminescence traces of the 520 nm (blue) and 540 nm (red) emission features from drop-cast films at 250 K (Methods). Data at both wavelengths is well represented by a bi-exponential decay (black) with a $\tau$=296 ps primary component. Reduced temperature increases the relative strength of the 540 nm emission feature. 250 K was chosen to make approximately equivalent the count rates at both features (inset).
    (\textbf{b}) Pump intensity-dependence of PL spectra from a drop-cast film (Methods) at T$\approx$300 K. Data has been normalized to the PL at 520 nm. 
    (\textbf{c}) Log-log plot of integrated PL strength ($I_\text{PL}$=$\int A(\lambda) d\lambda$) from ED (black) and MD (brown) contributions as a function of excitation intensity, $I_{\text{exc}}$. Lines are power law fits: $I_\text{PL}\propto I_\text{exc}^\alpha$ with nearly identical exponents for the ED ($\alpha$=1.17$\pm$0.05; black) and MD ($\alpha$=1.15$\pm$0.04; brown) emission.}\label{fig:TRPL_Pump}
\end{figure}

\noindent in \BAPbI \ exhibits an integrated magnitude [$\int A_\text{MD}(\lambda) d\lambda$] comparable to that of the ED, i.e.,
\ba{
\frac{\int A_\text{MD}(\lambda)d\lambda}
{\int A_\text{ED}(\lambda)d\lambda + \int A_\text{MD}(\lambda)d\lambda} 
\approx 0.16.
} As described below, such a strong multipolar PL contribution is highly unusual and particularly unexpected when we consider the overall fast dynamics of the system. In lanthanide ions, for instance, multipolar PL is observed due to long-lived ($\sim$100 $\mu$s) excited states associated with ED-forbidden recombination channels.\cite{sugano_multiplets_1970,dodson_magnetic_2012} Here, however, we observe sub-ns PL lifetimes for the main excitonic feature at 520 nm (Fig. \ref{fig:TRPL_Pump}a, blue circles). PL decay traces were measured at reduced temperature (250 K), at which phonon-induced population exchange between these two states is reduced\cite{wu_indirect_2019} (Methods). Still, PL traces of the isolated 540 nm feature (Methods) (Fig. 4a, red squares) show no discernible differences; both traces are well represented by a bi-exponential decay with a primary component lifetime of $\tau$$\approx$296 ps, consistent with previous reports. \cite{straus_direct_2016,xing_transcending_2017,cheng_highly_2018,proppe_synthetic_2018,kamminga_micropatterned_2018,becker_bright_2018} Unfortunately, this particular material shows a structural phase transition\cite{billing_synthesis_2007} below approximately 250 K, at which point the PL spectrum changes abruptly. While two distinct PL peaks separated by $\sim$90 meV are indeed visible in this low-temperature phase, their identity is ambiguous since

\noindent immersion oils cannot be used to perform energy-momentum spectroscopy at these low temperatures. We thus did not explore the detailed behavior of these features as a function of temperature. These two emission features also show nearly identical behavior under intensity-dependent PL. Throughout a very broad range of excitation intensities ($I_\text{exc}$=1-750 mW/cm$^2$) the spectral shape is observed to be nearly invariant (Fig. 4b; darker colored lines). Only at extremely high excitation intensities ($I_\text{exc}$$>$10$^3$ mW/cm$^2$; lighter colored lines) does the spectrum begin to vary, with the low-energy feature growing in relative strength over the high-energy feature.  Moreover, power-law fits of the decomposed spectra ($I_\text{PL}$$\propto$$I_\text{exc}^\alpha$) (Fig. \ref{fig:TRPL_Pump}c) reveal that both the ED contribution (black; $\alpha_\text{ED}$=1.17$\pm$0.05) and the multipolar contribution (brown; $\alpha_\text{MD}$=1.15$\pm$0.04) grow super-linearly ($\alpha$$>$1), but in parallel. Note that the data presented in Fig. \ref{fig:TRPL_Pump} was measured from poly-crystalline samples with random crystallite orientations in order to enhance the visibility of the 540 nm emission feature with conventional spectroscopy techniques (Methods), but similar conclusions are made from analogous measurements on spin-cast thin films (Supplementary Information S6; Supplementary Fig. 12). Together, these results suggest that both the ED and MD transitions are intrinsic in origin and are associated with excitonic states in thermal equilibrium, contrary to alternative speculations attributing the low-energy feature to bi-excitons ($\alpha$=2) or bound excitons ($\alpha$$<$1).

The \emph{absolute} radiative lifetimes can be estimated from the measured PL lifetime, quantum yield (PLQY), and normalized multipolar emission rates. Assuming that the measured PLQY of 0.4\% (Methods) is representative of both emission features, our data implies radiative rates of 1.4$\times$10$^7$ s$^{-1}$ and 1.9$\times$10$^6$ s$^{-1}$ for the ED and MD respectively, assuming the emission arises from identical exciton densities. (Note that this PLQY is comparable to previously reported values for this material.\cite{mitzi_synthesis_1996,blancon_extremely_2017,congreve_tunable_2017}) However, considering the 90 meV energy splitting, the observed low-energy (MD) transition rate could be enhanced relative to the high-energy (ED) transition due to the thermal occupation of excited states. If we assume two distinct emissive channels in thermal equilibrium,\cite{thorne_spectrophysics_1988} the intrinsic MD radiative rate is reduced by a factor of exp(90 meV/26 meV)$\approx$30, yielding an intrinsic radiative rate of 6.0$\times$10$^4$ s$^{-1}$. Even with this more conservative estimate, the inferred multipolar radiative rate is three orders of magnitude larger than any MD emission rate reported from atomic systems.\cite{dodson_magnetic_2012} There is, to our knowledge, no crystalline system to which this may be compared. The multipolar emission observed in \BAPbI \ is extraordinarily bright and challenges conventional understanding of multipolar light-matter interactions. 

\begin{figure}[H]
 	\includegraphics[width=3.5in]{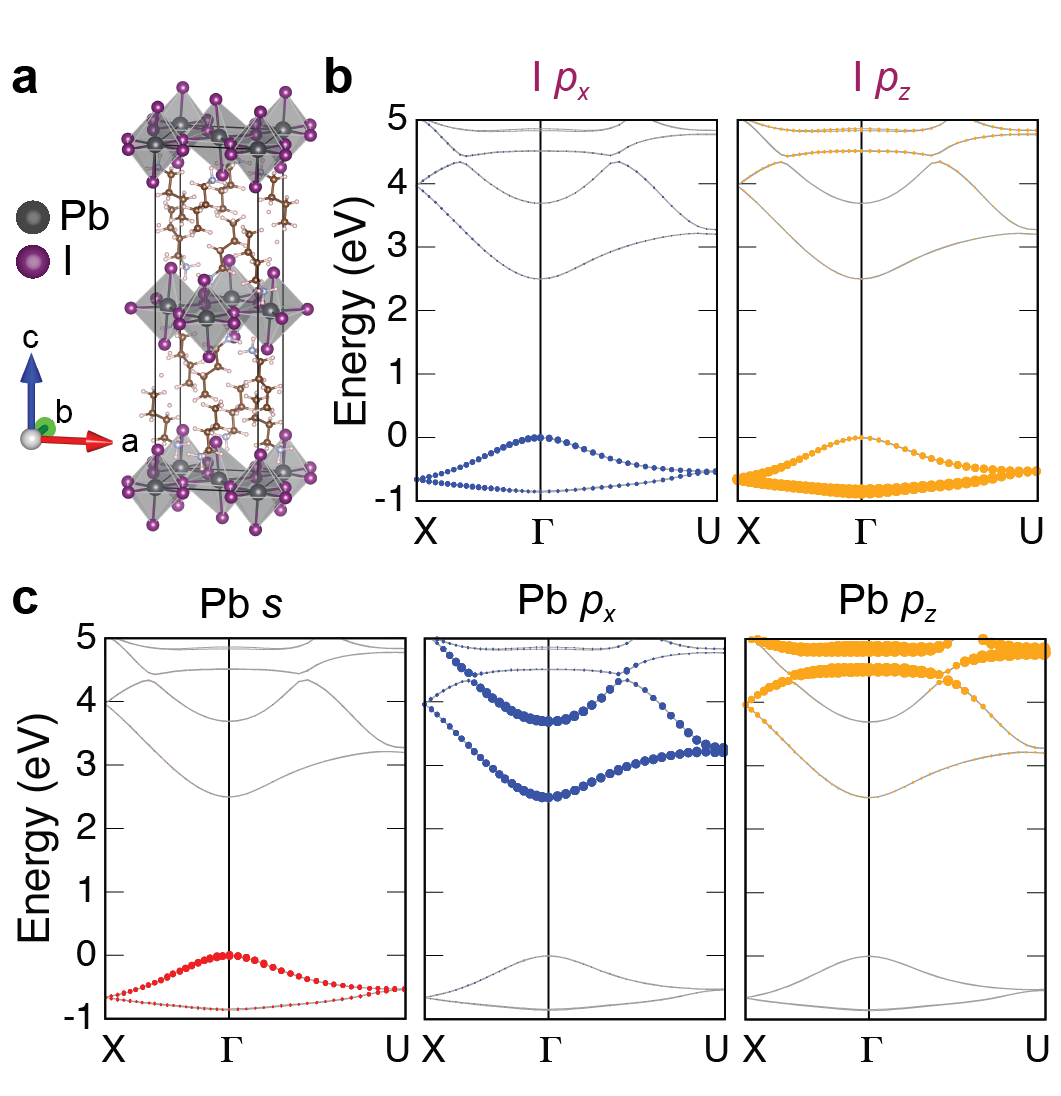}
	\caption{
	\textbf{\emph{Calculated band structure and band character of BA$_2$PbI$_4$.}}  
	\footnotesize
	(\textbf{a}) The \BAPbI \ unit cell used in DFT calculations comprises two distinct layers. Pb atoms are shown in black and I atoms are shown in purple. 
	(\textbf{b-c}) Computed band structures along the $\Gamma$-$X$ and $\Gamma$-$U$ directions of the reciprocal lattice, projected onto Pb and I orbitals with symmetries of $s$, $p_x$, $p_y$, and $p_z$. The relative weights are represented by the size of the circles at each momentum. 
	(\textbf{b}) I $p_x$ (left) and $p_z$ (right) contributions to the band structure. I (5)$p$ orbitals contribute primarily to the valence bands. The valence band maximum (VBM; energy set to 0 eV) predominantly derives from in-plane I (5)$p$ orbitals ($p_x$ and $p_y$). 
	(\textbf{c}) Pb $s$ (left), $p_x$ (center) and $p_z$ (right) contributions to the band structure.  The conduction band minimum (CBM) predominantly derives from in-plane Pb (6)$p$ orbitals ($p_x$ and $p_y$). The VBM also shows significant contributions from Pb (6)$s$ orbitals. In all cases, $p_y$ contributions are identical to $p_x$ contributions and are thus are not shown.}
	\label{fig:DFT}
\end{figure}

We now consider the electronic band structure as it relates to the origins of the multipolar PL signature. Density functional theory (DFT) calculations (Methods) of the \BAPbI \ electronic structure are summarized in Fig. \ref{fig:DFT}. At each wave vector, $\bf{k}$, along the $\Gamma$-X and $\Gamma$-U directions of the reciprocal lattice, we project the ground-state eigenfunctions onto a basis consisting of I (Fig. \ref{fig:DFT}b) and Pb (Fig. \ref{fig:DFT}c) orbitals with 

\noindent symmetries corresponding to $s$, $p_x$, $p_y$, and $p_z$ orbitals. The weight of the projection is represented by the size of the circle at each $\bf{k}$. The VBM (set to 0 eV in the band diagrams) comprises roughly equal contributions from I (5)$p$ orbitals and Pb (6)$s$ orbitals, in agreement with previous reports.\cite{umebayashi_electronic_2003,even_electronic_2012,filip_steric_2014,shirayama_optical_2016} (I and Pb $p_y$ contributions are not shown since they are nearly identical to $p_x$, due to the system's approximate $C_4$ symmetry.) Note, however, that the I $p_z$ contribution is significantly suppressed relative to the in-plane contributions. The CBM comprises nearly equal contributions of Pb (6)$p_x$ and (6)$p_y$ orbitals with an absent $p_z$ contribution, in agreement with ref. \citen{even_electronic_2012}.  (See Supplementary Fig. 13 for the complete set of band character projections and charge density plots.) The symmetries of the CBM and VBM states participating in optical transitions, however, are governed by the bonding character of the constituent atoms. Previous studies have assigned $\Gamma_6^-$ (``$p$-like") and $\Gamma_6^+$ (``$s$-like") symmetries to the CBM and VBM wavefunctions, $\ket{\psi_\text{CBM}}$ and $\ket{\psi_\text{VBM}}$, respectively.\cite{tanaka_electronic_2005,even_electronic_2012} In analogy with conventional semiconductors, these states may be qualitatively represented by $\ket{X}$ and $\ket{Y}$ (CBM), and $\ket{S}$ (VBM); the presence of a predominantly IP ED transition is described by a symmetry analysis of the ED matrix elements between CBM and VBM states: $\bra{S}x\ket{X}$$\approx$$\bra{S}y\ket{Y}$ and $\bra{S}z\ket{\psi_\text{CBM}}$$\approx$0.\cite{even_electronic_2012} The non-zero ED$_z$ contribution observed here and in previous experiments\cite{fieramosca_tunable_2018,guo_hyperbolic_2018} may arise from a non-negligible I $p_z$ contribution at the VBM and an I (5)$s$ contribution at the CBM (Supplementary Fig. 13). (While electron-hole correlations yield three distinct exciton levels with symmetries corresponding to the direct product of $\Gamma_6^-$$\otimes$$\Gamma_6^+$, the selection rules still reflect the underlying VBM and CBM states and these conclusions are thus unaltered.) This treatment has been used with apparent success to describe the low-temperature exciton spectrum observed in previous reports.\cite{ishihara_exciton_1989,tanaka_electronic_2005,even_electronic_2012,even_analysis_2014,even_importance_2013,giovanni_tunable_2016,yu_effective-mass_2016,fu_neutral_2017,becker_bright_2018} 

In contrast to the ED term, the second-order terms of the multipolar expansion (i.e., MD and EQ) connect states with \emph{equal} parity;\cite{bunker_molecular_2006} these multipolar transitions are strictly forbidden within the aforementioned treatment. Our experimental results thus suggest the presence of structural asymmetries that generate CBM and/or VBM states of mixed parity.\cite{sugano_multiplets_1970} Nonetheless, static DFT calculations reveal negligible static structural asymmetries, evidenced by both the spatial structure of the electronic wave functions\cite{even_electronic_2012} (Supplementary Fig. 13) and the absence of Rashba-like splitting of the energy bands (Fig. 5).\cite{stoumpos_ruddlesdenpopper_2016,zhang_first-principles_2018,zhang_three-dimensional_2018,frohna_inversion_2018} A surge of recent reports highlight the importance of dynamic symmetry-breaking mechanisms, e.g., significant thermally-induced polar lattice distortions of the relatively ``soft" ionic lattice. \cite{dow_toward_1972,kepenekian_rashba_2017,yaffe_local_2017,niesner_structural_2018,zheng_structural_2018,mckechnie_dynamic_2018, marronnier_influence_2019,wu_indirect_2019} For example,  ``dual emission" characteristics observed in HOIPs have been attributed to a distribution of indirect tail states caused by Rashba-like band splitting originating from large local polar distortions.\cite{wu_indirect_2019} Our observations are generally consistent with this report, but suggest a very different physical mechanism; rather than ED-mediated indirect transitions from Rashba split bands,\cite{kepenekian_rashba_2017,zhang_three-dimensional_2018,zhang_first-principles_2018,zheng_structural_2018,mckechnie_dynamic_2018,marronnier_influence_2019,wu_indirect_2019} relaxation may instead proceed from significantly perturbed states by direct transitions via the MD term. Such a dynamical origin likely also accounts for the relatively long Urbach-like tail observed for the MD transition.\cite{dow_toward_1972,wu_indirect_2019} Symmetry-based analysis of multipolar matrix elements is detailed in Supplementary Information S8. Qualitatively, MD$_z$ connects states with equivalent parity associated with in-plane extended orbitals. Since $\Gamma_6^-$ symmetry (associated with in-plane $p$-like orbitals) has already been assigned to the CBM, a possible explanation is the presence of a yet-to-be established odd-parity contribution to the VBM with strong in-plane character, deriving primarily from the contribution of I (5)$p$ orbitals (e.g., see Fig. \ref{fig:DFT}b). These underlying selection rules hold when the excitonic character of the excited state is considered,\cite{tanaka_electronic_2005,becker_bright_2018} and different electron-hole exchange energies arising from the presence of an additional term in the VBM (in combination with exchange and crystal field effects) may account for a portion of the observed 90 meV energy splitting. 

Note that this symmetry-based treatment does not provide estimates of the absolute magnitudes of the transitions, and the observed unconventionally fast MD transition impels deeper theoretical examination of atomic and electronic structure in 2D HOIPs. However, a brief examination of the predicted transition rates based on Fermi's Golden Rule suggests that such a fast MD transition is, in fact, not unreasonable in this material system. For simplicity, we describe ED and MD emitters in a homogeneous, isotropic, non-magnetic environment with refractive index $n$. Fermi's Golden Rule gives for the spontaneous emission rates from state $i$ to state $f$ through an MD ($A_\text{MD}$) channel:\cite{novotny_hecht_2006,taminiau_quantifying_2012}
\ba{
A_{\text{MD},i\rightarrow f} &= \frac{\omega^3 \epsilon_0^{1/2} \mu_0^{3/2}}{3 \hbar \pi} n^3 \left|\frac{q}{2 m c}\bra{f} \mathbf{L} + 2\mathbf{S} \ket{i} \right|^2
}

\noindent where $\omega$=$2\pi c/\lambda$, $c$ is the speed of light, $\epsilon_0$ ($\mu_0$) is the vacuum permittivity (permeability), $\hbar$ is Planck's constant, $q$ is the charge of the state in question, and $\mathbf{L}$ and $\mathbf{S}$ are the orbital and spin angular momentum operators. While ED rates in crystals are on the order of 10$^7$ s$^{-1}$, as observed here, MD and higher-order rates are often considered negligible by dimensional analysis arguments which typically invoke quantities relevant to isolated atoms ($r$$\sim$$a_0$, where $a_0$ is the Bohr radius) in homogeneous free-space ($n$=1), and thus assume bare electron masses ($m$=$m_0$). Detailed calculations for lanthanide ions\cite{dodson_magnetic_2012} give vacuum MD rates of 10-30 s$^{-1}$. Taking into account the refractive index ($n$$\approx$2.3) and the fact that the elementary excitations are states with effective mass $m^*$$\approx$0.1$m_0$ (rather than the bare electron mass, $m_0$), we estimate MD emission rates to be of order $n^3$/$(m^*/m_0)^2$$\approx$10$^3$ times greater than those in free space. This indeed brings us curiously close to the MD emission rate ($\sim$10$^4$ s$^{-1}$) observed here from HOIPs. Since these arguments are general, multipolar phenomena may be more prevalent than previously thought, perhaps existing in other 2D semiconductor systems or ionic crystals with strong spin-orbit coupling, unconventional band structures, heavy atoms (e.g., WSe$_2$.\cite{xia_transition_2017}) and significant lattice dynamics. That they have not readily been established may be a consequence of the rare usage of momentum- and polarization-resolved spectroscopies.

\section*{CONCLUSIONS}
Using energy-momentum spectroscopies, we establish that the low-energy sideband emission feature frequently observed in HOIPs exhibits an anomalous polarization and angle dependence. These unconventional radiation properties of the sideband luminescence have not yet been acknowledged. Consequently, the magnitude of these transitions has been largely underestimated, and interpretations of the underlying nature and significance of the emission have varied widely. Generally, this demonstrates the critical role of the experimental geometry on optical studies, particularly in the study of highly anisotropic material systems. Indeed, since such anisotropies are inherent to all novel 2D quantum materials, polarization and angle dependencies should be very carefully considered in future studies using conventional optical characterization techniques. These observations are critical for resolving discrepancies in the literature and may be exploited to optimize the functionality of optoelectronic devices, especially LEDs, based upon 2D HOIPs. For instance, the spectrum and dynamics of light emission may be manipulated by controlling this highly oblique multipolar emission with various photonic architectures.\cite{wiesmann_photonic_2009,li_quantifying_2014}

Exploiting the highly-oriented structure of 2D HOIPs, and taking into account these polarization-, wavelength-, and angle-dependent collection effects, we identify the sideband luminescence as arising from an oriented and exceptionally fast magnetic dipole optical transition. Although multipolar light-matter interactions have been demonstrated in III-V quantum dots, these higher-order interactions derive from mesoscale structuring\cite{andersen_strongly_2011} and strong electric field gradients in the vicinity of plasmonic structures.\cite{zurita-sanchez_multipolar_2002} That is, the multipolar interactions are characteristic of electronic envelope functions enforced by the structure of the quantum dot rather than the symmetries of the Bloch functions associated with the bulk material. In contrast, the presence of a multipolar transition in 2D HOIPs is evidently inherent to the material system and is understood on the basis of symmetries of the Bloch functions used to describe both the valence and conduction bands. Accounting for these extrinsic collection effects, we show that the MD radiative rate is at least three orders of magnitude larger than multipolar emission rates previously identified in atomic systems;\cite{taminiau_quantifying_2012,dodson_magnetic_2012,karaveli_time-resolved_2013} there is yet no other crystalline system to which these rates may be compared. This discovery of bright MD luminescence, coupled with first-principles considerations of radiation rates, suggest that multipolar phenomena may be more prevalent than previously thought, particularly in materials with, e.g., strong confinement, large spin-orbit coupling, mixed parity states, or dynamic structural distortions.

\section*{Methods}
\footnotesize 
\textbf{Samples.} 
Solutions of \BAPbILong \ were prepared according to the procedures described in ref. \citen{venkatesan_charge-carrier_2018} to concentrations of 150 mg/mL in an N$_2$-filled glovebox. Thin films were dynamically spin-cast (solution deposited while spinning) in an N$_2$-filled glovebox on 0.180 mm fused silica substrates at 8000 RPM for 60 seconds and subsequently thermally annealed at 70$^\circ$C for 30 minutes, producing films with thicknesses of $\sim$61 nm. Film thicknesses were measured with atomic force microscopy (AFM) (Asylum MFP-3D) in tapping mode using a tip-scratch method and measuring over the resulting groove. For variable angle spectroscopic ellipsometry studies, thin films were prepared identically, upon Si substrates. All samples were kept in a nitrogen-vacated dry box immediately following film preparation until the time of optical measurements. Mechanically exfoliated flakes (Supplementary Figs. 5-7) were produced by the ``scotch tape method" from single crystals grown according to the methods described in ref. \citen{ledee_fast_2017} in an N$2$-filled glovebox. Flake thicknesses were determined by AFM in tapping mode after optical measurements were made. Single crystals of \BAMAPbILong \ with n=1 and n=2 (Supplementary Fig. 3) were grown from solution directly on plasma-cleaned fused silica substrates by drop depositing $\sim$7 $\mu$L of solution (prepared as described above) and allowing the solvent to evaporate at elevated temperatures ($\sim$45$^\circ$C). Samples were subsequently thermally annealed at 100$^\circ$C for 10 minutes. Crystals were identified visually by means of an inverted microscope with magnification varying from 10$\times$ to 100$\times$, and phase purity was confirmed by analyzing the PL spectra (Supplementary Fig. 5). Thin films of (C$_4$H$_9$NH$_3$)$_2$PbBr$_4$ (butylammonium lead bromide; Supplementary Fig. 6) were prepared from \BAPbILong \ thin films by a vapor exchange method according to procedures specified in Supplementary Information S4. Thin films of (C$_8$H$_{17}$NH$_3$)$_2$PbI$_4$ (octylammonium lead iodide) and (C$_{12}$H$_{25}$NH$_3$)$_2$PbI$_4$ (dodecylammonium lead iodide) (Supplementary Fig. 7) were prepared according to the procedures described above with the substitution of butylammonium for octylammonium and dodecylammonium, respectively. For TRPL and pump power-dependence studies (Fig. \ref{fig:TRPL_Pump}), samples were prepared by the drop-casting method (described above) without initial plasma cleaning of substrates. This strategically produced thick films with mixed crystallite orientations and significant scattering texture by minimizing wetting of the substrate, thereby enhancing the visibility of the 540 nm emission without the use of high-NA imaging techniques. 

\noindent \textbf{GIWAXS.}
Grazing-incidence wide-angle X-ray scattering (GIWAXS) experiments were conducted at the Advanced Light Source at beamline 7.3.3. The energy of the incident beam was at 10 keV and a Pilatus 2M area detector was used. The X-ray scattering data were taken at incidence angles of 0.3$^\circ$ and samples were kept in a helium environment during X-ray exposure to minimize air scattering and sample degradation. The collected images were processed using GIXSGUI,\cite{jiang_gixsgui:_2015} and the beam center and the sample-to-detector distance were calibrated using silver behenate (AgBe).

\noindent \textbf{Optical constants.} 
Uniaxial optical constants for thin films of \BAPbILong \ were determined by a combination of momentum-resolved reflectometry\cite{decrescent_model-blind_2016} and variable angle spectroscopic ellipsometry.  See Supplementary Information S1 for complete details.  

\noindent \textbf{Energy-momentum spectroscopy.} 
Thin film samples were illuminated by a 405 nm light-emitting diode (LED) (ThorLabs M405L3). The incident light was filtered by a 405 nm short-pass filter and reflected by a 415 nm dichroic mirror to remove overlap with the emission wavelengths. The emitted PL was transmitted through the same dichroic as well as a 417 nm long-pass filter. The energy and $y$-momentum, $k_y$, of the light was then measured by imaging the objective (Nikon CFI Plan Fluor 100$\times$ Oil, 1.3 NA) BFP to the entrance slit of an imaging spectrometer (Princeton Instruments IsoPlane SCT320 with Princeton Instruments PIXIS 1024BRX). An analyzing polarizer was applied to the image such that light along the $k_y$-axis was either $p$-polarized (vertical polarizer alignment) or $s$-polarized (horizontal polarizer alignment). For the data presented in Figs. \ref{fig:mPL} and \ref{fig:decomposition}, measurements were performed in ambient conditions with illumination intensities below $\sim$1 mW/cm$^2$. The exposed films showed no visible signs of photo-degradation. The resulting energy-momentum spectra were decomposed into various multipolar contributions using methods described elsewhere\cite{taminiau_quantifying_2012,schuller_orientation_2013} assuming a uniform distribution of dipole/multipole locations along the $z$-axis. Film thicknesses were taken from AFM (see ``Samples"). Film optical constants were determined from the same films using methods described completely in Supplementary Information S1. For photo-degradation studies (Supplementary Fig. 11), measurements were performed in both ambient and inert (N$_2$) conditions, and illumination intensities were swept between $\sim$5 mW/cm$^2$ and 100 mW/cm$^2$. No visible signs of photo-degradation were observed for samples exposed in inert conditions. Integration times for all measurements were on the order of 10-1000 s, depending upon illumination intensity. For energy-momentum spectroscopy of exfoliated flakes, after sample preparation (see ``Samples"), a heavy flow of nitrogen gas was applied to the sample.  The nitrogen flow persisted for the duration of the optical measurements. Single crystals and platelets were identified by eye under a 100$\times$ oil-immersion objective.  Single crystals were illuminated by focusing a 405 nm fiber-coupled laser to the image plane at the rear port of the microscope.  At the level of the sample, the focused spot had a width of $\sim$5 $\mu$m.  Total illumination intensities were kept below $\sim$0.06 W/cm$^2$. Integration times were on the order of 100-1000 s. The exposed crystals showed no significant signs of optical degradation. All energy-momentum spectroscopy measurements and pump-dependent PL measurements were performed at room temperature. 

\noindent \textbf{TRPL.}
Time-correlated single-photon counting measurements (Fig. \ref{fig:TRPL_Pump}a) were performed in the UCSB Optical Characterization Facility. Approximately 100 fs excitation pulses with wavelength 400 nm were generated by doubling the fundamental frequency of fs Ti:Sapphire laser (Spectraphysics Tsunami) pulses in a commercial optical harmonic generator (Inrad). The laser repetition rate was reduced to 2 MHz by a home-made acousto-optical pulse picker in order to avoid saturation of the chromophore. The system is equipped with thermoelectrically-cooled single-photon counting avalanche photodiode (Micro Photon Devices) and electronics board (Becker \& Hickl SPC-630) and has instrument response time ~30 ps. The fluorescence signal was dispersed in Acton Research SPC-500 monochromator after passing through long-pass filter to remove the pump wavelength. In addition to the time-resolved detector, the monochromator is equipped with a CCD camera (Roper Scientific PIXIS-400) allowing for monitoring of the time-averaged fluorescence spectrum. To maximize the 540 nm PL signal, films were produced by the drop-casting method as described in ``Samples". Samples were mounted within an evacuated cryostat maintained at $\sim$0.5$\times$10$^{-5}$ mBar to avoid photo-degradation. Samples were pumped at 400 nm with $\approx$0.025 $\mu$J/cm$^2$ per pulse with a 2 MHz repetition rate. PL was collected in a reflection geometry with the pump and collection optics at 90$^\circ$ with respect to each other and approximately 45$^\circ$ with respect to the sample normal. The data shown in Fig. \ref{fig:TRPL_Pump}a was acquired at 250 K in order to minimize thermally assisted excitation transfer between the states. Analogous measurements presented in the Supporting Information were performed on spin-cast thin films by collecting the PL at high angles (with respect to the sample interfaces) through a linear polarizer in an s-polarized configuration.  

\noindent \textbf{Pump-dependent PL.}
For pump power-dependence studies as presented in Fig. 4b-c, films were prepared according to the drop-casting method as described in ``Samples". Regions with reoriented crystals or sufficient scattering texture were identified visually. The reorientation and/or scattering texture allowed the 540 nm emission to be detected with a 10$\times$ 0.3 NA (Nikon CFI Plan Fluor). Samples were illuminated by a loosely focused spot ($\sim$100 $\mu$m beam waste) in inert (N$_2$) conditions according to the methods described for single crystals above. Illumination intensities were swept between $\sim$1 mW/cm$^2$ and 1 W/cm$^2$. No signs of photo-degradation were observed for pump powers below $\sim$750 mW/cm$^2$. Integration times for all measurements were on the order of 0.1-250 s, depending upon illumination intensity. Spectra were decomposed to a sum of two (asymmetric) Gaussian distributions with stretched tails and fixed widths. The centers of the distributions were held at 522 nm and 541 nm. Only the amplitudes of the two components were fit at each illumination intensity.

\noindent \textbf{PLQY.} 
PLQY of spin-cast thin films was measured using a Horiba spectrometer (FluoroMax 4) and integrating sphere (Horiba Quanta-$\varphi$) with 400 nm wavelength excitation. To improve PL counts, samples were spin-cast at 1500 rpm resulting in films of $\sim$300 nm thickness. The photon absorption rate was determined by scanning over the excitation line of the attenuated beam (nominal OD=3, measured area fill correction of 803) with the sample in place and comparing to a non-absorbing reference (blank quartz substrate). A 405 nm short-pass filter (Semrock) was in place during excitation and emission measurements, and a 405 nm long-pass filter (Semrock) was in place for emission measurements only. A 1\% correction factor was applied to account for long-pass transmission efficiencies. We note that the 540 nm emission was clearly visible in integrating sphere PL measurements, as expected since PL is collected over all angles. 

\noindent \textbf{Band structure.}
First-principles calculations were performed based on DFT as implemented in the Vienna Ab initio Simulation Package (VASP)\cite{kresse_efficient_1996}. We used projector augmented wave (PAW)\cite{blochl_projector_1994} pseudopotentials with a plane-wave energy cutoff of 500 eV. A $\Gamma$-centered $3\times 3\times 1$ {\bf k}-point grid was used for sampling the first Brillouin zone. The Heyd-Scuseria-Ernzerhof (HSE)\cite{heyd_hybrid_2003} hybrid functional was employed for the exchange-correlation interactions. A mixing parameter of 0.45 was used for the HSE functional including spin-orbit coupling to give a consistent bandgap with experiment. The atomic positions and lattice constants were fully relaxed; the optimized lattice constants were 8.79, 8.80 and 29.07 {\AA}, respectively.

\section*{Acknowledgements}
Energy-momentum spectroscopy measurements and analyses were supported by the National Science Foundation (DMR-1454260 and OIA-1538893) and by the Air Force Office of Scientific Research (Grant No. FA9550-16-1-0393).
Time- and intensity-resolved optical measurements and analyses, and studies of exfoliated single crystals comprising varying organic spacers were supported as part of Quantum Materials for Energy Efficient Neuromorphic Computing, an Energy Frontier Research Center funded by the U.S. Department of Energy (DOE), Office of Science, Basic Energy Sciences (BES), under Award \# DE-SC0019273.
Materials synthesis and structural characterization were supported by the U.S. Department of Energy, Office of Science, Basic Energy Sciences, under Award DE-SC-0012541. 
PLQY measurements were performed at the MRL Shared Experimental Facilities are supported by the MRSEC Program of the NSF under Award No. DMR 1720256; a member of the NSF-funded Materials Research Facilities Network (www.mrfn.org). 
Portions of the work were performed at the Advanced Light Source, supported by the Director, Office of Science, Office of Basic Energy Sciences, of the U.S. Department of Energy under Contract No. DEAC02-05CH11231. 
A portion of this work was performed in the UCSB Nanofabrication Facility. 
The research reported here also made use of the shared facilities of the UCSB MRSEC (National Science Foundation DMR 1720256), a member of the Materials Research Facilities Network (www.mrfn.org). 
R.M.K. gratefully acknowledges the National Defense Science and Engineering Graduate fellowship for financial support.
X.Z. was supported by the U.S. Department of Energy (DOE), Office of Science, Basic Energy Sciences (BES) under Award No. DE-SC0010689. Computational resources were provided by the National Energy Research Scientific Computing Center, a DOE Office of Science User Facility supported by the Office of Science of the U.S. Department of Energy under Contract No. DE-AC02-05CH11231.
We thank Chris G. Van de Walle and Michael E. Flatte for helpful discussions.

\section*{Author contributions}
R.A.D., M.L.C., and J.A.S. conceived of the experiment. 
R.A.D. and W.L. performed optical measurements. R.A.D. performed electromagnetic calculations and analyzed the optical data with R.Z. and J.A.S. 
R.A.D., N.R.V., C.J.D., R.M.K. synthesized samples. 
N.R.V., C.J.D. and R.M.K. performed and analyzed GIWAXS measurements with M.L.C. 
X.Z. performed numerical band structure calculations. 
J.A.S. supervised the project. 
All authors discussed the results and commented on the manuscript.

\section*{Additional information}
\textbf{Supplementary Information} accompanies this paper. \\
\textbf{Data availability:} The data that support the findings of this study are available from the corresponding author upon reasonable request.
\textbf{}

\end{multicols}

\newpage

\bibliographystyle{bibstyle_NatMat}
\bibliography{man_bib_mainDoc}

\end{document}


\newcommand{\printtitle}{\textbf{Supplementary Information: Bright magnetic dipole radiation from two-dimensional lead-halide perovskites}}
\newcommand{\printauthors}{
Ryan A. DeCrescent,\textsuperscript{1} 
Naveen R. Venkatesan,\textsuperscript{2} 
Clayton J. Dahlman,\textsuperscript{2}  
Rhys M. Kennard,\textsuperscript{2}  
Xie Zhang,\textsuperscript{2}
Wenhao Li,\textsuperscript{3}
Xinhong Du,\textsuperscript{4}
Michael L. Chabinyc,\textsuperscript{2}  
Rashid Zia\textsuperscript{3}  
and
Jon A. Schuller\textsuperscript{*,4} 
}
\newcommand{\printaffiliations}{
\textsuperscript{1}\emph{Department of Physics, University of California Santa Barbara, Santa Barbara, CA 93106, USA}
\\
\textsuperscript{2}\emph{Materials Department, University of California Santa Barbara, Santa Barbara, CA 93106, USA}
\\
\textsuperscript{3}\emph{Brown University, School of Engineering, Providence, Rhode Island 02912, USA.}
\\
\textsuperscript{4}\emph{Department of Electrical and Computer Engineering, University of California, Santa Barbara, California 93106, United States}
}
\newcommand{\printcorrespondence}{*Correspondence should be addressed to J.A.S. (email: jonschuller@ece.ucsb.edu)}

\begin{center}
\printtitle \\[10pt]
\printauthors \\[10pt]
\printaffiliations \\[10pt]
\printcorrespondence
\end{center}

\section*{Table of Contents}
\noindent Section S1: Optical constants of \BAPbILong \ thin films \\
\noindent Section S2: Calculating normalized intrinsic multipolar emission rates \\
\noindent Section S3: Polarization-dependent shoulder emission in related layered lead-halide perovskites \\
\noindent Section S4: Photo-degradation effects upon the multipolar emission energy \\
\noindent Section S5: Radiation patterns from oriented multipoles \\
\noindent Section S6: TRPL and pump power-dependence of multipolar decomposition in thin films \\
\noindent Section S7: Numerical band-structure calculations and charge densities \\
\noindent Section S8: Matrix elements of multipolar transitions

\newpage

\section{Optical constants of \BAPbILong \ thin films}\label{Sect:Methods_Constants}
\subsection{Momentum-resolved reflectometry}\label{Sect:Methods_Constants_mR}
Single-shot momentum-resolved reflectometry (mR) \cite{decrescent_model-blind_2016} was performed at discrete wavelengths spanning 500-800~nm by uniformly illuminating the back focal plane (BFP) of an oil-immersion (Nikon 100x, 1.3NA, Plan Fluor) objective with a coherent white light source (NKT Photonics SuperK EXTREME) passed through a multi-wavelength filter (NKT Photonics SuperK SELECT, Vis-nIR and nIR1 configurations) \cite{decrescent_model-blind_2016}.  This illuminates the thin film samples, with discrete selected wavelengths, at angles ranging from normal incidence ($k_{||}$=0, $\theta$=0$^\circ$) to approximately $k_{||}$/$k_0$=1.3 ($\theta$$\approx$64$^\circ$) from within the substrate, well beyond the critical angle for total internal reflection.  The specular reflection was collected with the same objective, and the momentum ($k_{||}$) distribution of the specular reflection (i.e., as imaged by the objective's BFP) was projected to the entrance slit of an imaging spectrometer (Princeton Instruments IsoPlane SCT 320).  For $s$-($p$-)polarized measurements, both the incident and reflected beams were polarized so that light along the $\hat{y}$-momentum axis was $\hat{x}$-($\hat{y}$-)polarized.  The momentum-resolved reflection spectra were normalized against a high quality fused silica reference (equivalent to sample substrates) at each wavelength to account for polarization- and momentum-dependent collection efficiencies.  

\hspace{3em} The ordinary in-plane (IP) complex refractive index, $\tilde{n}_o$, was fit at each wavelength by numerically finding the minimum of the least-squares error of the measured $s$-polarized reflectance with respect to a three-layer uniaxial Fresnel model.  Using $\tilde{n}_o$ as an input, the extraordinary out-of-plane (OP) complex refractive index, $\tilde{n}_e$, was determined similarly by comparing $p$-polarized reflectance with respect to a three-layer uniaxial Fresnel model.  (See ref. \cite{decrescent_model-blind_2016} for complete details.)  Advanced beyond the basic technique as presented in ref. \cite{decrescent_model-blind_2016}, the theoretical reflection curves were ``corrected" in a number of ways: (1) theory curves were convoluted with a Gaussian smearing function in $k_{||}$-space to account for finite BFP imaging resolution (point spread function); (2) we quantitatively account for surface roughness by assuming the reflection profile to be an incoherent sum of reflecting domains with total thicknesses given by a Gaussian distribution about the mean thickness, with a Gaussian width determined by the RMS surface roughness.  (Both the mean and RMS thicknesses were taken from AFM.)  We believe this to be a very good approximation for films with thicknesses ($\approx$50 nm) much smaller than the length scale over which the thickness varies maximally ($\sim$$\mu$m). Based upon our own experience, we find that both the real ($n_o$) and imaginary parts ($k_o$) of the ordinary complex refractive index ($\tilde{n}_o$) are determined with high precision and accuracy.  While the imaginary part ($k_e$) of the extraordinary complex refractive index ($\tilde{n}_e$) is determined with similar precision, we find the results for the real part of the extraordinary complex refractive index admit large uncertainty and are often underestimated.  Total illumination intensities were kept below $\approx$1 mW/cm$^2$.

\subsection{Variable angle spectroscopic ellipsometry (VASE)}\label{Sect:Methods_Constants_VASE}
VASE data was collected with a J.A. Woollam model M2000DI for angles between 45$^\circ$ and 75$^\circ$, and wavelengths between 300 nm and 1500 nm.  Each sample was measured at four different rotational orientations (i.e., by rotating the sample by 90$^\circ$ between each measurement).  A bare Si sample (`substrate') from the same wafer was also measured as a ``reference" sample to eliminate fit parameters associated with the substrate and native oxide layer (`layer \#1').  The sub-bandgap (transparent, and thus real) uniaxial refractive index of each sample was determined by modeling the film as a uniaxial layer (`layer \#2') with independent Cauchy distributions for both in-plane (ordinary) and out-of-plane (extraordinary) directions.  A non-linear (`exponent'=3.0) graded interface layer was assumed to account for surface roughness, with a roughness layer thickness determined uniquely by statistical analyses of AFM measurements.  The optical constants of the graded interface layer were modeled according to the Bruggeman effective medium approximation as a combination of the thin film (layer \#2) and vacuum.  Film thicknesses for each VASE measurement were allowed to vary during the fitting procedure within the ranges observed by AFM.  Averages and standard deviations of results were estimated from the four measurements.  We find that this procedure gives a simple and reliable estimate of optical anisotropies in thin film systems, with the minimal set of fit parameters.  All VASE measurements were performed in ambient conditions.

\subsection{Unifying mR and VASE results}
To obtain reasonable estimates of all optical constants, continuously united over the whole spectral range ($\lambda$=495-1500 nm), we first parameterize $\tilde{n}_o$ results to agree with mR (in the range 500-800 nm) and with VASE (in the transparent region, approximately 900-1500 nm, depending on the value of $m$).  We adapt this parameterization to agree with $k_e$ (as determined by mR) and with $n_e$ (as determined by VASE in the transparent region of the films).  The parameterization also serves to ensure Kramers-Kronig (KK) consistency.  The simplest KK-consistent model within this wavelength range offering the best fits for \BAPbI \ consisted of a Tauc-Lorentz oscillator (representing band-edge absorption) and a Gaussian oscillator (representing the single sub-bandgap excitonic feature).  The model for $\tilde{n}_e$ was assumed to relate to $\tilde{n}_o$ solely by a variation in amplitude of the oscillators and the high-frequency permittivity ($\epsilon_\infty$), and the center energy of the excitonic absorption feature was allowed to vary between ordinary and extraordinary dimensions to account for the obvious transverse-longitudinal splitting observed from mR. To account for the uncertainty in $n_e$ in the infrared (as determined by multiple VASE measurements), the parameterized extraordinary refractive index was fit to both the ``maximum" (average + 2$\sigma$) and minimum (average - 2$\sigma$) estimates of the infrared VASE results.  In performing this procedure, even without ``formal" ellipsometry fitting over the whole spectral range, we find the models generate very good representations of the measured ellipsometric parameters $\Delta(\lambda)$ and $\Psi(\lambda)$. The uniaxial optical constants used in this study are presented in Fig. \ref{Fig:Constants}. A complete study of optical anisotropies in \BAMAPbI\ with $n$=1 through $n$=4 and \MAPbI\ will be presented in a future publication.

\begin{figure}[H]
\begin{center}
  \includegraphics[width=0.5\linewidth]{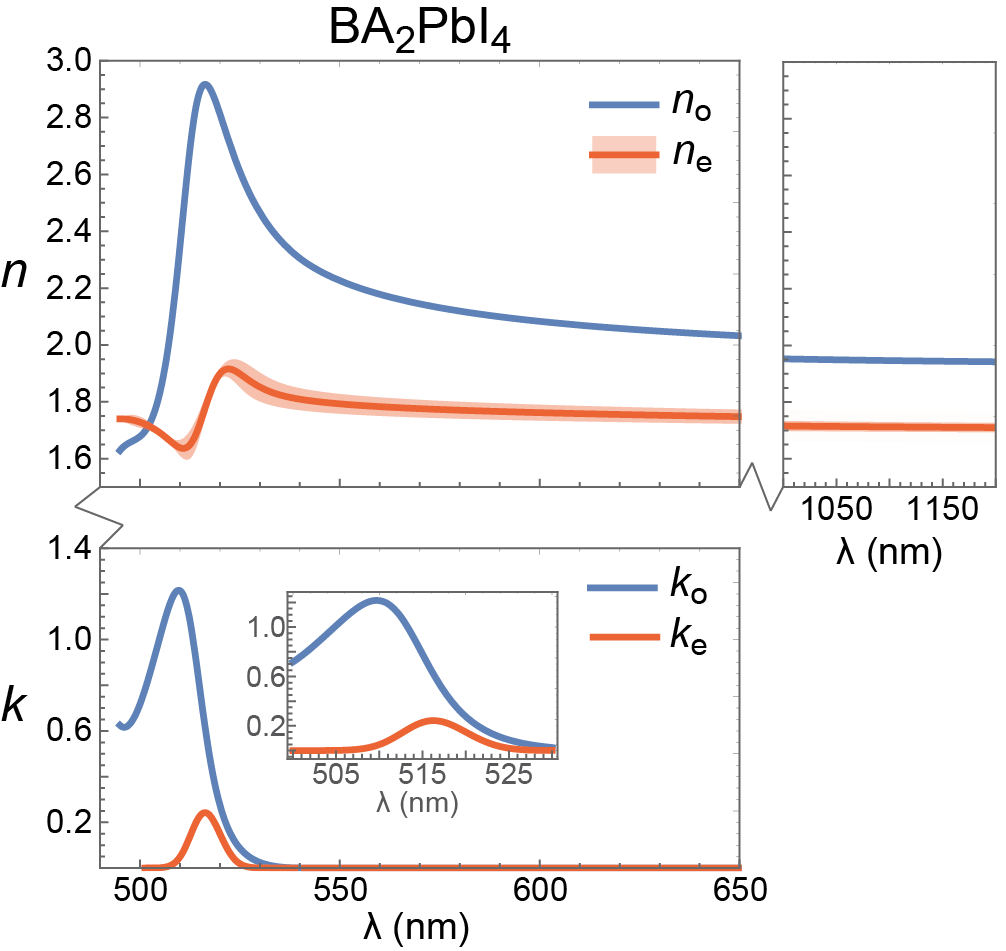}
  \caption[]{\textbf{\emph{Complex uniaxial refractive index of BA$_2$PbI$_4$.}}
  The (top) real and (bottom) imaginary component of the uniaxial complex refractive index. The in-plane (ordinary) component is shown in blue. The out-of-plane (extraordinary) component is shown in red. Estimated 95\% confidence intervals for $n_e$ are represented by the red shaded region. The inset shows the excitonic absorption peaks over a narrower range of wavelengths. The approximate 5:1 ratio of IP:OP absorption strengths is consistent with measurements on single-crystals \cite{ishihara_exciton_1989} indicating that spin-cast thin films of \BAPbI \ are structurally similar to single crystals with PbI$_4$ planes parallel to the substrate.}
  \label{Fig:Constants}
  \end{center}
\end{figure}

\newpage
\begin{figure}[H]
\begin{center}
  \includegraphics[width=0.8\linewidth]{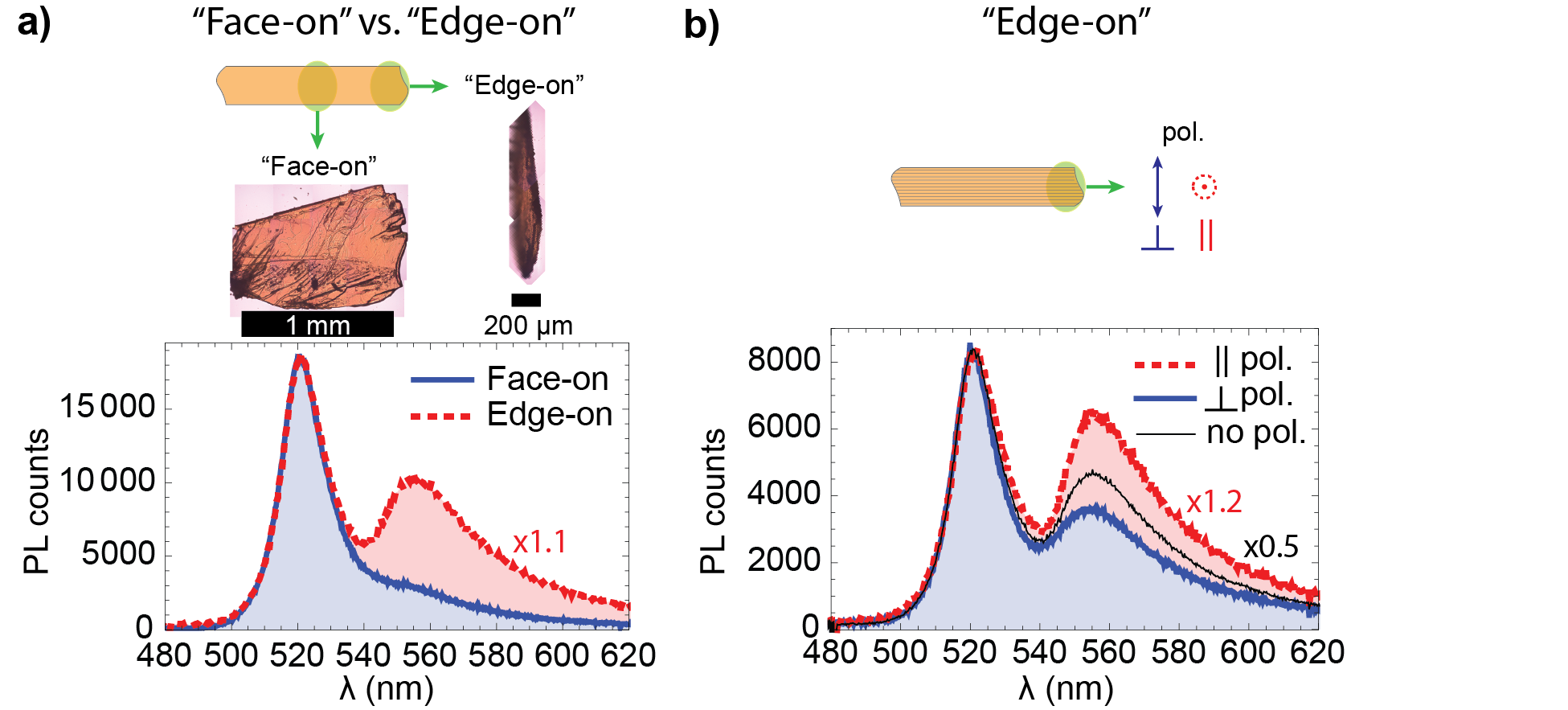}
  \caption[]{\textbf{\emph{Impact of experimental geometry on measured PL.} }
  PL spectra from a millimeter-scale single crystal of \BAPbI \ (grown using methods specified in ref. \cite{stoumpos_ruddlesdenpopper_2016}) as collected with a 0.75 NA objective (Nikon CFI Plan Fluor 40$\times$ 0.75 NA) and different experimental geometries with respect to the crystal.
  (\textbf{a}) Unpolarized PL spectra from the (solid blue) face of the crystal and from the (dashed red) edge of the crystal.
  (\textbf{b}) Polarized PL spectra from the edge of the crystal. Electric field polarization conventions are specified in the schematic. Solid blue; perpendicular ($\perp$) polarizer alignment (electric field perpendicular to lead-iodide layers). Dashed red; parallel ($||$) polarizer alignment (electric field parallel to lead-iodide layers). The reference unpolarized spectrum is shown as a thin black line. Spectra have been normalized to the 520 nm peak (scale factors specified by colored text in each plot). Optical images of the crystal face and edge are shown in panel (a). Respective scale bars are specified for each visual image. The low-energy peak is observed in the perpendicular alignment because the edge of the crystal naturally has significant scattering texture, as seen in the visual image.}
  \label{Fig:Constants}
  \end{center}
\end{figure}

\newpage
\section{Calculating normalized intrinsic multipolar emission rates}\label{Sect:LDOS_Multipolar}

In energy-momentum spectroscopy, the $s$- or $p$-polarized PL counts, $N^{s,p}(\om,\kpar)$, measured at a given frequency ($\om$=2$\pi c/\lambda$, where $\lambda$ is the free-space wavelength) and a given (conserved) in-plane electromagnetic wave momentum, $\bf{k}_{||}$=$k_x\hat{x}$+$k_y\hat{y}$, is given by

\ba{
N^{s,p}(\om,\kpar)
= C_\text{exp} \left[ 
\Gamma_{ED_x}^{s,p}(\om,\kpar) 
+\Gamma_{ED_y}^{s,p}(\om,\kpar) 
+\Gamma_{ED_z}^{s,p}(\om,\kpar) 
+\Gamma_{MD_z}^{s,p}(\om,\kpar) 
+\Gamma_{EQ_{xy}}^{s,p}(\om,\kpar)
+\Gamma_{EQ_{x'y'}}^{s,p}(\om,\kpar)
\right]
}

Here, the $\Gamma_{M_o}$ correspond to the photon emission rates of a multipole, $M$ (i.e., ED, MD, EQ), with an orientation (or ``configuration", for EQs) $o$, positioned in the thin film (an inhomogeneous environment), and $C_\text{exp}$ is a constant accounting for experimental collection efficiencies. As in the manuscript, we focus on in-plane (IP) and out-of-plane (OP) oriented EDs, OP-oriented MDs, and the two degenerate IP transverse EQs (EQ$_{xy}$ and EQ$_{x'y'}$). According to refs \cite{taminiau_quantifying_2012,schuller_orientation_2013}, the multipolar emission rates, $\Gamma_{M_o}^{s,p}(\om,\kpar)$, are related to the intrinsic emission rates, $A_{M_o}$, in a bulk (infinite) homogeneous medium of refractive index $n$ by the normalized local density of optical states (LDOS), $\rho_{M_o}^{s,p}$, according to 
\ba{
\Gamma_{M_o}^{s,p}(\omega,\kpar) = A_{M_o}(\omega) \tilde{\rho}_{M_o}^{s,p}(\omega,\kpar)
}

giving
\ba{
N^{s,p}(\om,\kpar)
&= C_\text{exp} \left[ 
A_{ED_x}(\om)\tilde{\rho}_{ED_x}^{s,p}(\om,\kpar) 
+A_{ED_y}(\om)\tilde{\rho}_{ED_y}^{s,p}(\om,\kpar) 
+A_{ED_z}(\om)\tilde{\rho}_{ED_z}^{s,p}(\om,\kpar) \right. \nonumber \\
& \hspace{1cm} \left. 
+A_{MD_z}(\om)\tilde{\rho}_{MD_z}^{s,p}(\om,\kpar) 
+A_{EQ_{xy}}(\om)\tilde{\rho}_{EQ_{xy}}^{s,p}(\om,\kpar)
+A_{EQ_{x'y'}}(\om)\tilde{\rho}_{EQ_{x'y'}}^{s,p}(\om,\kpar)
\right]
}

We allow the intrinsic multipolar emission rates, $A_{M_o}$, to vary with orientation since it is possible, in general, to have different oriented species \cite{schuller_orientation_2013}. However, we will assume $A_{ED_x}$=$A_{ED_y}$ and $A_{EQ_{xy}}$=$A_{EQ_{x'y'}}$, as necessary for a rotationally invariant system. By absorption-emission reciprocity, the enhanced (or suppressed) emission rate into each mode can be determined by the enhanced (or suppressed) absorption rate for illumination by that mode:
\ba{
\tilde{\rho}_{M_o,\text{emit}}^{s,p}\equiv\frac{\Gamma_{M_o}^{s,p}}{A_{M_o}}
=\tilde{\rho}_{M_o,\text{absorb}}^{s,p}\equiv\frac{\alpha_{M_o}^{s,p}}{B_{M_o}}
}

in which we have defined two multipolar absorption rates analogous to the emission rates: $\alpha_{M_o}^{s,p}(\om,\kpar)$ is the absorption rate in the thin film when illuminated by an $s$- or $p$-polarized plane wave of frequency $\omega$ and IP wave vector $\kpar$; $B_{M_o}(\om)$ is the total absorption rate in a homogeneous environment (i.e., integrated over all electromagnetic modes with wave vector $k$=$|\mathbf{k}|$=$n\omega/c$). Particularly, 

\ba{
\tilde{\rho}_{ED_i}^{s,p}(\om,\kpar) &= 
\frac{|\mu_{ED,i}E_i^{s,p}(\om,\kpar)|^2}
{\sum_{s,p}\int d^2k|\mu_{ED,i} E_{0,i}^{s,p}(\om,\kpar)|^2} =
\frac{|\tilde{E}_i^{s,p}(\om,\kpar)|^2}
{\sum_{s,p}\int d^2k|\hat{e}_i^{s,p}(\kpar)|^2}
\label{Eqn:ED_LDOS}
\\
\tilde{\rho}_{MD_i}^{s,p}(\om,\kpar) &= 
\frac{|\mu_{MD,i}\tilde{B}_i^{s,p}(\om,\kpar)|^2}
{\sum_{s,p}\int d^2k|\mu_{MD,i} B_{0,i}^{s,p}(\om,\kpar)|^2} =
\frac{|\tilde{B}_i^{s,p}(\om,\kpar)|^2}
{\sum_{s,p}\int d^2k|\hat{b}_i^{s,p}(\kpar)|^2}
\label{Eqn:MD_LDOS}
\\
\tilde{\rho}_{EQ_{xy}}^{s,p}(\om,\kpar) &= 
\frac{|Q_{xy}\partial_x E_y^{s,p}(\om,\kpar) + Q_{yx}\partial_y E_x^{s,p}(\om,\kpar)|^2}
{\sum_{s,p}\int d^2k|Q_{xy} \partial_x E_{y,0}^{s,p}(\om,\kpar) + Q_{yx} \partial_y E_{x,0}^{x,p}(\om,\kpar)|^2} 
=
\frac{|\partial_x \tilde{E}_y^{s,p}(\om,\kpar) + \partial_y E_x^{s,p}(\om,\kpar)|^2}
{\sum_{s,p}\int d^2k|\partial_x \hat{e}_y^{s,p}(\kpar) + \partial_y \hat{e}_x^{x,p}(\kpar)|^2} 
\label{Eqn:EQ_LDOS}
}

Here, $\tilde{E}_i^{s,p}$=$E_i^{s,p}/E_0$ ($\tilde{B}_z^s$=$B_z^s/B_0$) is the electric (magnetic) field amplitude, projected along the $\hat{i}$ direction, at the position of the emitter, normalized by the field amplitude, $E_0$ ($B_0$), of the incident wave. The fields $\hat{e}_i^{s,p}$ and $\hat{b}_i^{s,p}$ are the frequency-independent unit-amplitude electric and magnetic field vectors projected along the Cartesian $\hat{i}$ direction ($\hat{x}_i$): $\hat{e}_i^{s,p}$=$(\hat{x}_i\cdot\vec{E}_0^{s,p})/E_0$. The EQ coefficients cancel because $Q_{yx}$=$Q_{xy}$. Notice that $s$-polarized radiation from ED emitters ($\Gamma_{ED,i}^s$) is uniquely determined by up to a constant scale factor since only IP EDs contribute to this emission, i.e., $\tilde{\rho}_{ED,z}^s$=0.
\\

\subsection{ED LDOS Denominators}
The numerators for the ED LDOS are readily available in the literature \cite{schuller_orientation_2013} and can be expressed by matrix methods \cite{azzam_ellipsometry_1977}. We evaluate the denominators for the three unique dipole orientations using cylindrical coordinates in the variable $\kpar$ ($|\kpar|$=$\kp$). In projecting the spherical integration surface (radius $k$=$|\mathbf{k}|$=$n\omega/c$) to the circular integration surface over the variable $\kpar$ (0$\leq$$|\kpar|$$\leq$$k$), we must include a factor of 1/$\cos\theta$=$k/(k^2-k_{||}^2)^{1/2}$ in the integrand to ensure that all waves carry equivalent power in $\hat{z}$:
\ba{
i&=x: \hspace{0.25cm} 
2\iint \frac{\kp d\kp d\phi}{\cos\theta}(|\hat{e}_x^s(\kpar)|^2 + |\hat{e}_x^p(\kpar)|^2) =
2\iint \frac{\kp d\kp d\phi}{\cos\theta}
(\sin^2\phi +\cos^2\theta \cos^2\phi) = \frac{8\pi k^2}3
\\
i&=y: \hspace{0.25cm} 
2\iint \frac{\kp d\kp d\phi}{\cos\theta}(|\hat{e}_y^s(\kpar)|^2 + |\hat{e}_y^p(\kpar)|^2) =
2\iint \frac{\kp d\kp d\phi}{\cos\theta}
(\cos^2\phi +\cos^2\theta \sin^2\phi) = \frac{8\pi k^2}3
\\
i&=z: \hspace{0.25cm} 
2\iint \frac{\kp d\kp d\phi}{\cos\theta}(|\hat{e}_z^s(\kpar)|^2 
+|\hat{e}_z^p(\kpar)|^2) =
2\iint \frac{\kp d\kp d\phi}{\cos\theta}
(0 + \sin^2\theta) = \frac{8\pi k^2}3
}

where $\cos\phi$=$k_x/\kp$ ($0$$<$$\phi$$<$$2\pi$), and the factor of 2 is included for integration over $0$$<$$\theta$$<$$\pi/2$ to account for the full 4$\pi$ steradians. Equality of the three integrals is expected on the grounds that the reference space is isotropic. Identical results are obtained by performing integration over spherical coordinates ($\theta,\phi$) and eliminating the factor of $1/\cos\theta$ from the integrand.

\subsection{MD LDOS}
The quantities $|\tilde{B}_i^{s,p}|^2$ in the slab (thin film) can be calculated analytically \cite{taminiau_quantifying_2012}. Here we use S-Matrix methods to directly relate these terms to $|\tilde{E}_i^{s,p}|^2$. Note that, for a plane wave propagating with a wave vector $\mathbf{k}$, the electric and magnetic fields are related by $\mathbf{k} \times \mathbf{E}$=$\omega \mathbf{B}$.  Component by component,
\ba{
B_x = \frac1\omega \left( k_y E_z - k_z E_y \right) \\
B_y = \frac1\omega \left( - k_x E_z + k_z E_x \right) \\
B_z = \frac1\omega \left( k_x E_y - k_y E_x \right) \\
}

To express the fields corresponding to $s$- and $p$-polarized plane waves, take $\kpar$=$k_x$$\hat{x}$=$\kp\hat{x}$. With this geometry, $\mathbf{E}^s$=$E_y \hat{y}$ ($\mathbf{B}^s$=$B_x\hat{x} + B_z\hat{z}$) and $\mathbf{E}^p$=$E_x \hat{x} + E_z \hat{z}$ ($\mathbf{B}^p$=$B_y\hat{y}$). In the slab (thin film), the total field is comprised of both `+' (upward propagating) and `-' (downward propagating) components. The law of reflection gives $(k_{z}^{s,p})^-$=$-(k_{z}^{s,p})^+$$\equiv$$- k_{z}^{s,p}$. 
\ba{
p\text{-pol.}: 
\hspace{0.5cm} 
B_{x}^p &= B_{z}^p = 0 \\
B_{y}^p &= (B_{y}^p)^+ + (B_{y}^p)^- \nonumber \\
&= \frac1\omega \{ [- k_x (E_{z}^p)^+ + k_{z}^p (E_{x}^p)^+] + [- k_x (E_{z}^p)^- + (k_{z}^p)^- (E_{x}^p)^-] \} \nonumber \\
&= - \frac1\omega \{ k_x [(E_{z}^p)^+ + (E_{z}^p)^-] - k_{z}^p [(E_{x}^p)^+ - (E_{x}^p)^-]\} \nonumber \\
&= - \frac1\omega \{ k_x E_{z}^p - k_{z}^p [(E_{x}^p)^+ - (E_{x}^p)^-]\} \nonumber \\
&= - \frac1\omega \{ \kp E_{z}^p - k_{z}^p [(E_{||}^p)^+ - (E_{||}^p)^-]\} \\
B_z^p &= 0 \\
\vspace{0.5cm}
s\text{-pol.}: 
\hspace{0.5cm} 
B_{x}^s &= (B_{x}^s)^+ + (B_{x}^s)^- \nonumber \\
&= \frac1\omega [ -(k_{z}^s)^+ (E_{y}^s)^+ - (k_{z}^s)^- (E_{y}^s)^- ] \nonumber \\
&= - \frac1\omega k_{z}^s [(E_{y}^s)^+ - (E_{y}^s)^-] \nonumber  \\
&= - \frac1\omega k_{z}^s [(E_{||}^s)^+ - (E_{||}^s)^-] \\
B_{y}^s &= 0 \\
B_{z}^s &= (B_{z}^s)^+ + (B_{z}^s)^- \nonumber \\
&= \frac1\omega [\kp (E_{||}^s)^+ + \kp (E_{||}^s)^- ] \\
&= \frac1\omega \kp E_{||}^s \nonumber 
}

where we have defined $E_{||}^s$=$E_y^s$ and $E_{||}^p$=$E_x^p$ as in in-plane component of the $s$- and $p$-polarized fields for the illumination configuration with $\kpar$=$\kp\hat{x}$ as a matter of convenience. Evaluating the normalized field magnitude squared (numerators in Eqn. \ref{Eqn:MD_LDOS}) with the relation $B_0$=$(k/\omega)E_0$ for each Cartesian component,

\ba{
| \tilde{B}_{x,z}^p|^2 &= 0 \\
|\tilde{B}_y^p|^2
&= 
\frac1{k^2} \left| \kp \tilde{E}_z^p - k_z^p [(\tilde{E}_{||}^p)^+ - (\tilde{E}_{||}^p)^-] \right|^2 \\
| \tilde{B}_x^s|^2 
&= \frac{|k_z^s|^2}{k^2} \left| (\tilde{E}_{||}^s)^+ - (\tilde{E}_{||}^s)^- \right|^2   \\
| \tilde{B}_y^s|^2 &= 0 \\
| \tilde{B}_z^s|^2
&= \frac{k_{||}^2}{k^2} | \tilde{E}_{||}^s|^2 
}

The denominators of Eqn. \ref{Eqn:MD_LDOS} are simply related to the denominators of Eqn. \ref{Eqn:ED_LDOS} by acknowledging the following relations:
\ba{
\hat{b}_x^p(\kpar) = \hat{e}_x^s(\kpar)
\hspace{0.5cm}& 
\hat{b}_x^s(\kpar) = -\hat{e}_x^p(\kpar) \\
\hat{b}_y^p(\kpar) = \hat{e}_y^s(\kpar) 
\hspace{0.5cm}&
\hat{b}_y^s(\kpar) = -\hat{e}_y^p(\kpar) \\
\hat{b}_z^p(\kpar) = \hat{e}_z^s(\kpar) = 0
\hspace{0.5cm}&
\hat{b}_z^s(\kpar) = -\hat{e}_z^p(\kpar)
}

Therefore, the denominators of Eqn. \ref{Eqn:MD_LDOS} evaluate to the same values as those of Eqn. \ref{Eqn:ED_LDOS}:
\ba{
\sum_{s,p}\int d^2k|\hat{b}_i^{s,p}(\kpar)|^2 = \frac{8\pi k^2}{3} \hspace{0.5cm} \text{for all $i$}
}

which is again just the statement that MDs of all orientations in a homogenous and isotropic environment emit at equal rates.

\subsection{EQ LDOS}
Of interest here are the two distinct EQ terms (EQ$_{xy}$ and EQ$_{x'y'}$). The denominators are equivalent in both cases due to the azimuthal symmetry of the reference system:

\ba{
B_{EQ_{xy}} &=
2\iint \frac{\kp d\kp d\phi}{\cos\theta}
(|\partial_x \hat{e}_y^s(\kpar) + \partial_y \hat{e}_x^s(\kpar)|^2
+|\partial_x \hat{e}_y^p(\kpar) + \partial_y \hat{e}_x^p(\kpar)|^2) \nonumber \\
&= 
2\iint \frac{\kp d\kp d\phi}{\cos\theta}
(|i k_x \hat{e}_y^s(\kpar) + i k_y \hat{e}_x^s(\kpar)|^2
+|i k_x \hat{e}_y^p(\kpar) + i k_y \hat{e}_x^p(\kpar)|^2) \nonumber \\
&=
2\iint \frac{\kp d\kp d\phi}{\cos\theta}
\kp^2
(|\cos\phi \hat{e}_y^s(\kpar) + \sin\phi \hat{e}_x^s(\kpar)|^2
+|\cos\phi \hat{e}_y^p(\kpar) + \sin\phi \hat{e}_x^p(\kpar)|^2) \nonumber \\
&=
2\iint \frac{\kp^3 d\kp d\phi}{\cos\theta}
(|-\cos^2\phi + \sin^2\phi|^2
+|-\cos\phi \cos\theta \sin\phi  - \sin\phi \cos\theta \cos\phi|^2) \nonumber \\
&=\frac{4\pi k^4}3
}

where we have again evaluated the integral in cylindrical coordinates with the same angle conventions as in the previous section. Other EQ configurations can be analyzed analogously. \\

For EQ$_{xy}$, we work out the numerator with $\kpar$=$k_{||}\hat{x}$, noting that the only $x$ dependence in the problem is contained in the phase factor $\exp(i k_{||} x)$: 
\ba{
|\partial_x \tilde{E}_y^s(\om,\kp)|^2
= |i k_{||} \tilde{E}_y^s(\om,\kp)|^2
= k_{||}^2 |\tilde{E}_y^s(\om,\kp)|^2
}

For EQ$_{x'y'}$, which is equivalent to EQ$_{xy}$ rotated by $\pi/4$ radians in-plane, we consider again $\kpar$=$\kp\hat{x}$ and define rotated coordinates $x'$=$(x+y)/\sqrt2$ and $y'$=$(x-y)/\sqrt2$ which are ``square'' with EQ$_{x'y'}$ and write the driving term:
\ba{
|\partial_{x'} \tilde{E}_{y'}^p(\om,\kp) + \partial_{y'} \tilde{E}_{x'}^p(\om,\kp)|^2
&=
\left|(\frac{1}{\sqrt2}\partial_x + \frac{1}{\sqrt2}\partial_y) \tilde{E}_{y'}^p(\om,\kp) + (-\frac{1}{\sqrt2}\partial_y + \frac{1}{\sqrt2}\partial_x)\tilde{E}_{x'}^p(\om,\kp) \right|^2 \nonumber \\
&= 
\frac{k_{||}^2}2 |\tilde{E}_{y'}^p(\om,\kp)+\tilde{E}_{x'}^p(\om,\kp)|^2 \nonumber \\
&=
\frac{k_{||}^2}2 \left| \frac{2}{\sqrt2}\tilde{E}_{x}^p(\om,\kp) \right|^2 \nonumber \\
&=
k_{||}^2 |\tilde{E}_{x}^p(\om,\kp)|^2 
}

(Note that this calculation is equivalent to considering EQ$_{xy}$ with an incident wave propagating with $\kpar$=$\kp(\hat{x}+\hat{y})/\sqrt2$.)
Finally,
\ba{
\tilde{\rho}_{EQ_{xy}}^s(\om,\kp) = \frac{3}{4\pi k^4} k_{||}^2|\tilde{E}_{||}^s(\om,\kp)|^2 \\
\tilde{\rho}_{EQ_{x'y'}}^p(\om,\kp) = \frac{3}{4\pi k^4} k_{||}^2|\tilde{E}_{||}^p(\om,\kp)|^2
}

where we have defined $\tilde{E}_{||}^s$$\equiv$$\tilde{E}_y^s$ and $\tilde{E}_{||}^p$$\equiv$$\tilde{E}_x^p$ as the in-plane component of the $s$- and $p$-polarized fields for the illumination configuration with $\kpar$=$\kp \hat{x}$ as a matter of convenience. Note that $\tilde{\rho}^s_{EQ_{xy}}$ is functionally equivalent to $\tilde{\rho}^s_{MD_z}$.

\subsection{Producing $x$- and $y$-polarized BFP images}
With the above expressions, $x$- and $y$-polarized 2D BFP radiation profiles for any emitting species (e.g., ED, MD, EQ) are produced by projecting the $s$- and $p$-polarized expressions onto a Cartesian basis. Explicitly:

\ba{
x\text{-pol.:} \hspace{0.5cm}
\rho^x(k_x,k_y) &=
\sin^2\phi \ \rho^s(\kp) + \cos^2\phi \ \rho^p(\kp)
= \frac{k_y^2}{k_x^2+k_y^2} \rho^s\left(\sqrt{k_x^2+k_y^2}\right) + \frac{k_x^2}{k_x^2+k_y^2} \ \rho^p\left(\sqrt{k_x^2+k_y^2}\right) \\
y\text{-pol.:} \hspace{0.5cm}
\rho^y(k_x,k_y) &=
\cos^2\phi \ \rho^s(\kp) + \sin^2\phi \ \rho^p(\kp)
= \frac{k_x^2}{k_x^2+k_y^2} \rho^s\left(\sqrt{k_x^2+k_y^2}\right) + \frac{k_y^2}{k_x^2+k_y^2} \rho^p\left(\sqrt{k_x^2+k_y^2}\right)
}

\newpage
\section{Polarization-dependent shoulder emission in related layered lead-halide perovskites}\label{Sect:OtherSystems}
\subsection{Preparation of butylammonium lead-bromide thin film samples}
Bromide Ruddlesden-Popper perovskites were prepared via halide substitution of iodide Ruddlesden-Popper perovskites \cite{solis-ibarra_post-synthetic_2015}. Briefly, the vapor substitution setup of Solis-Ibarra et al. was reproduced with the following modifications: All nitrogen gas lines were purged for 5 min. 1.25 mL of liquid bromine was poured into a 50 mL round-bottom flask and allowed to equilibrate for 2 min. The bromine gas stream was combined with a nitrogen gas stream, leading into a second 50 mL round-bottom flask (the sample chamber). The thin films were suspended over the chamber for 1 min., enabling full reaction with bromine. This procedure is known to preserve the crystallinity and morphology of the initial perovskite sample, and was thus preferred over spin-coating bromide Ruddlesden-Popper perovskites from bromide-containing precursor solutions for this particular study.

\begin{figure}[H]{}
	\centering
	\begin{minipage}{0.53\textwidth}
		\centering
		\includegraphics[width=\linewidth]{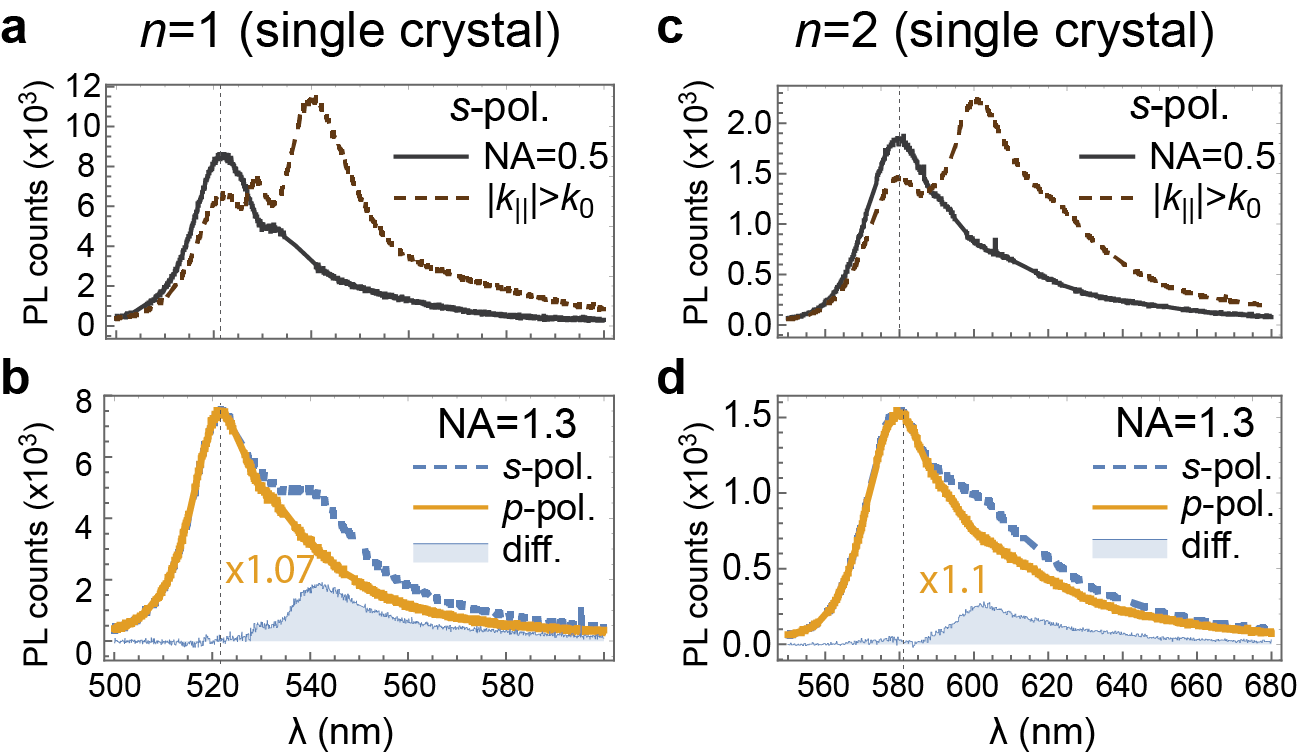}
		\caption{
		\textbf{\emph{Multipolar emission in single crystals of \BAMAPbI \ with n=1-2.}}  
		Polarization- and momentum-dependent PL spectra of single crystals of \BAMAPbI \ with (\textbf{a},\textbf{b}) $n$=1 and (\textbf{c},\textbf{d}) $n$=2. In all cases, a secondary strong emission peak at high momenta ($|k_{||}|$$>$$k_0$) arises in the $s$-polarized spectrum at an energy $\approx$90 meV below the primary ED-mediated exciton emission peak (vertical dashed lines).
		(\textbf{a}) $s$-polarized PL spectra collected at $|k_{||}|$$<$0.5$k_0$ (NA=0.5; solid) and $|k_{||}|$$>$$k_0$ (dashed).  A small peak at 528 nm 	(533 nm) in the $|k_{||}|$$>$$k_0$ ($|k_{||}|$$<$0.5$k_0$) spectrum is due to a Fabry-Perot-type oscillation (crystals are of $\sim$150 nm thickness) and is not associated with a distinct transition.
		(\textbf{b}) Momentum-integrated (NA=1.3) $s$- (blue dashed) and $p$-polarized (orange solid) PL spectra. The $p$-polarized spectrum has been normalized to give a peak height at 520 nm equal to that of the $s$-polarized spectrum (normalization factor printed in plot). The difference spectrum is represented by the blue filled region.
		(\textbf{c}) Same as (a), but for single crystals with $n$=2.
		(\textbf{d}) Same as (b), but for single crystals with $n$=2.
		A dashed vertical line in all subpanels indicates the primary exciton transition of the respective material.
		}
		\label{fig:single_xtal}
	\end{minipage}\hfill
	\begin{minipage}{0.45\textwidth}
		\centering
		\includegraphics[width=\linewidth]{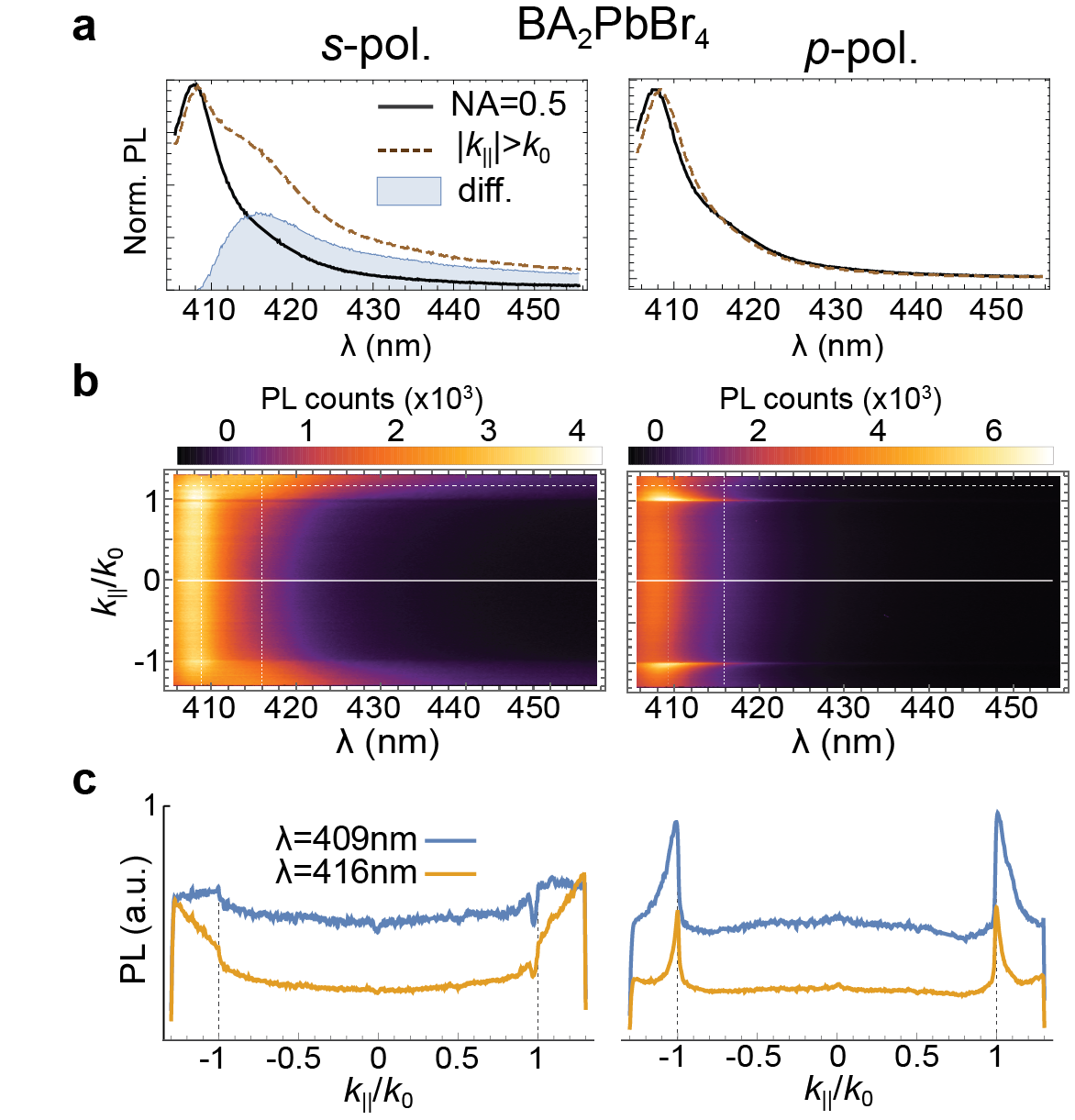}
		\caption{
		\textbf{\emph{Multipolar emission in butylammonium lead bromide.}}
		Thin films of butylammonium lead bromide (BA$_2$PbBr$_4$) show a polarization- and momentum-dependent emission signature very similar to that observed in thin films of BA$_2$PbI$_4$ (Fig. 1 of the manuscript).   
		(\textbf{a}) $s$-Polarized (left) and $p$-polarized (right) PL spectra collected at $|k_{||}|$$<$$0.5k_0$ (NA=0.5; solid) and $|k_{||}|$$>$$k_0$ (dashed). The ED-mediated exciton peak is at 409 nm, as observed in both $s$- and $p$-polarized spectra.  The $s$-polarized high-$k_{||}$ spectrum shows a distinct low-energy emission feature centered at 416 nm. PL traces are normalized to be equivalent at the primary exciton peak (409 nm). The difference between $|\kp|$$>$$k_0$ $|\kp|$$<$$0.5k_0$ $s$-polarized spectra is shown by the blue filled region.
		(\textbf{b}) The energy-momentum spectra from which the PL spectra of (a) were taken.
		(\textbf{c}) $s$- (left) and $p$-polarized (right) momentum-space emission profiles taken at $\lambda$=409 nm (blue) and $\lambda$=416 nm (orange).
	}
	\label{fig:BAPbB_mPL}
	\end{minipage}
\end{figure}

\begin{figure}[H]{}
\centering
	\includegraphics[width=0.9\linewidth]{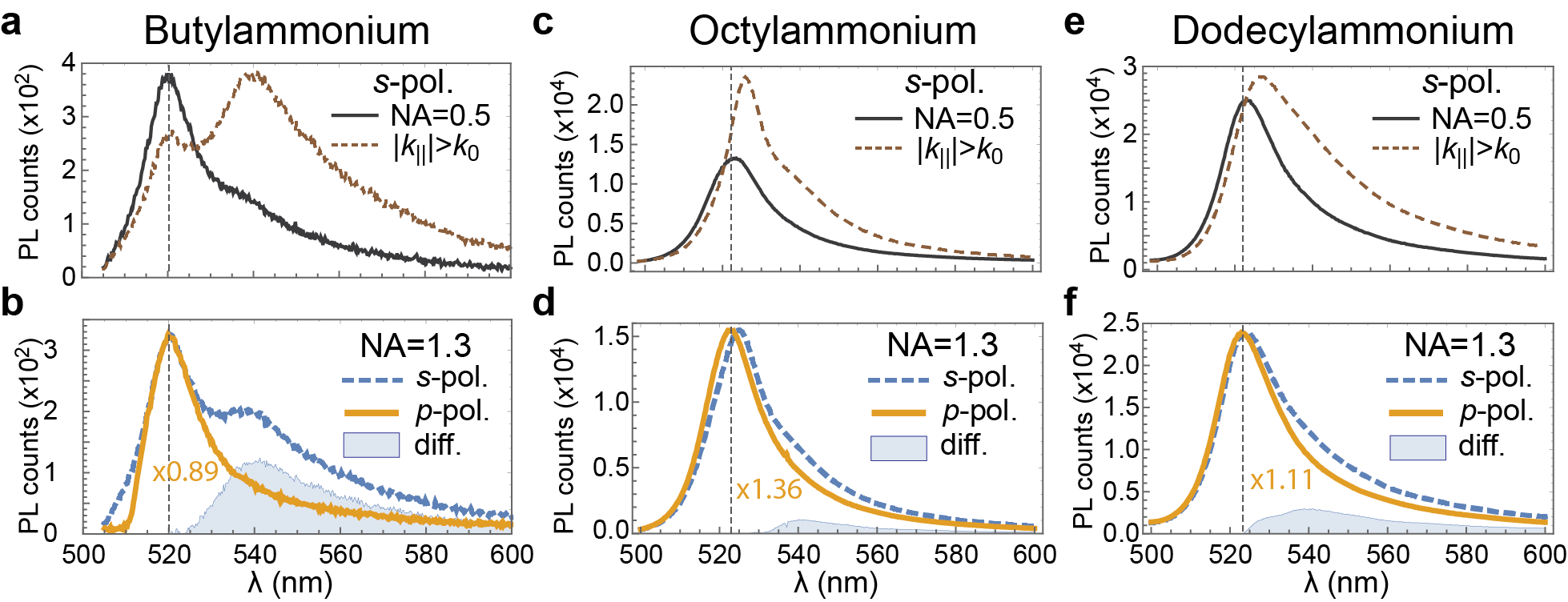}
	\caption{
	\textbf{\emph{Polarization-dependent PL in various alkylammonium-PbI$_4$ spin-cast thin films.}} \\
	Polarization- and momentum-dependent PL spectra of thin films of A$_2$PbI$_4$ with (\textbf{a},\textbf{b}) `A'=butylammonium (BA), (\textbf{c},\textbf{d}) `A'=octylammonium (OA), and (\textbf{e},\textbf{f}) `A'=dodecylmmonium (DDA). In all cases, a secondary emission peak or shoulder is observed at high momenta ($|k_{||}|$$>$$k_0$) in the $s$-polarized spectrum. The energy difference between the primary ED-mediated exciton emission (vertical dashed lines) and the inferred shoulder maximum is $\approx$20 nm ($\approx$90 meV) in all cases.
	(\textbf{a}) $s$-polarized PL spectra of BA$_2$PbI$_4$ collected at $|k_{||}|$$<$0.5$k_0$ (NA=0.5; solid lines) and $|k_{||}|$$>$$k_0$ (dashed lines; same as Fig. 1a of manuscript).
	(\textbf{b}) Momentum-integrated (NA=1.3) $s$- (blue dashed) and $p$-polarized (orange solid) PL spectra.  The $p$-polarized spectrum has been normalized to give a peak height equal to that of the $s$-polarized spectrum (normalization factor printed in plot). The difference of two normalized spectra is shown as a blue shaded region.
	(\textbf{c}) Same as (a), but for OA$_2$PbI$_4$.
	(\textbf{d}) Same as (b), but for OA$_2$PbI$_4$.  
	(\textbf{e}) Same as (a), but for DDA$_2$PbI$_4$.
	(\textbf{f}) Same as (b), but for DDA$_2$PbI$_4$.  
	A dashed vertical line in all subpanels indicates the primary exciton transition of the respective material.
	}
	\label{fig:OAPbI_mPL}
\end{figure}

\newpage
\begin{figure}[H]{}
	\centering
	\begin{minipage}{0.45\textwidth}
		\centering
		\includegraphics[width=\linewidth]{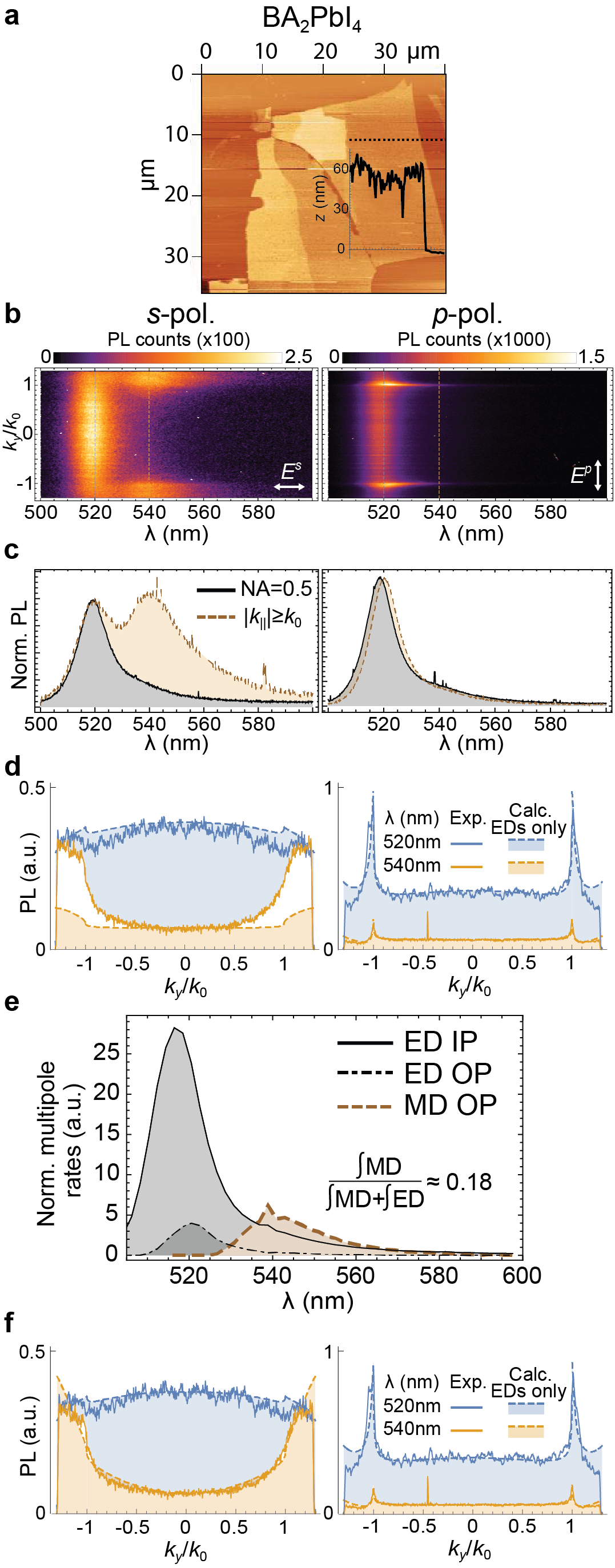}
		\caption{
		\textbf{\emph{Summarized analysis of exfoliated single crystals of BA$_2$PbI$_4$.}}
		\footnotesize
		(\textbf{a}) AFM image and horizontal trace of a flake produced by mechanical exfoliation of a single crystal.
		(\textbf{b}) Energy-momentum spectra. Left: $s$-polarization. Right: $p$-polarization.
		(\textbf{c}) Normalized PL traces at $|\kp|$$>$$k_0$ (dashed) and $|\kp|$$<$$0.5k_0$.
		(\textbf{d}) $\kp$ linecuts of energy-momentum spectra at 520 nm (blue) and 540 nm (orange).
		(\textbf{e}) Multipolar decomposition of energy-momentum spectra into oriented EDs (black) and OP MDs (brown).
		(\textbf{f}) Experimental (solid) and theoretical (dashed) $\kp$ linecuts at 520 nm (blue) and 540 nm (orange) including both oriented EDs and MDs. Optical constants were assumed to be identical to those of thin films.
		} 
		\label{fig:flake_BA}
	\end{minipage}\hfill
	\begin{minipage}{0.45\textwidth}
		\centering
		\includegraphics[width=\linewidth]{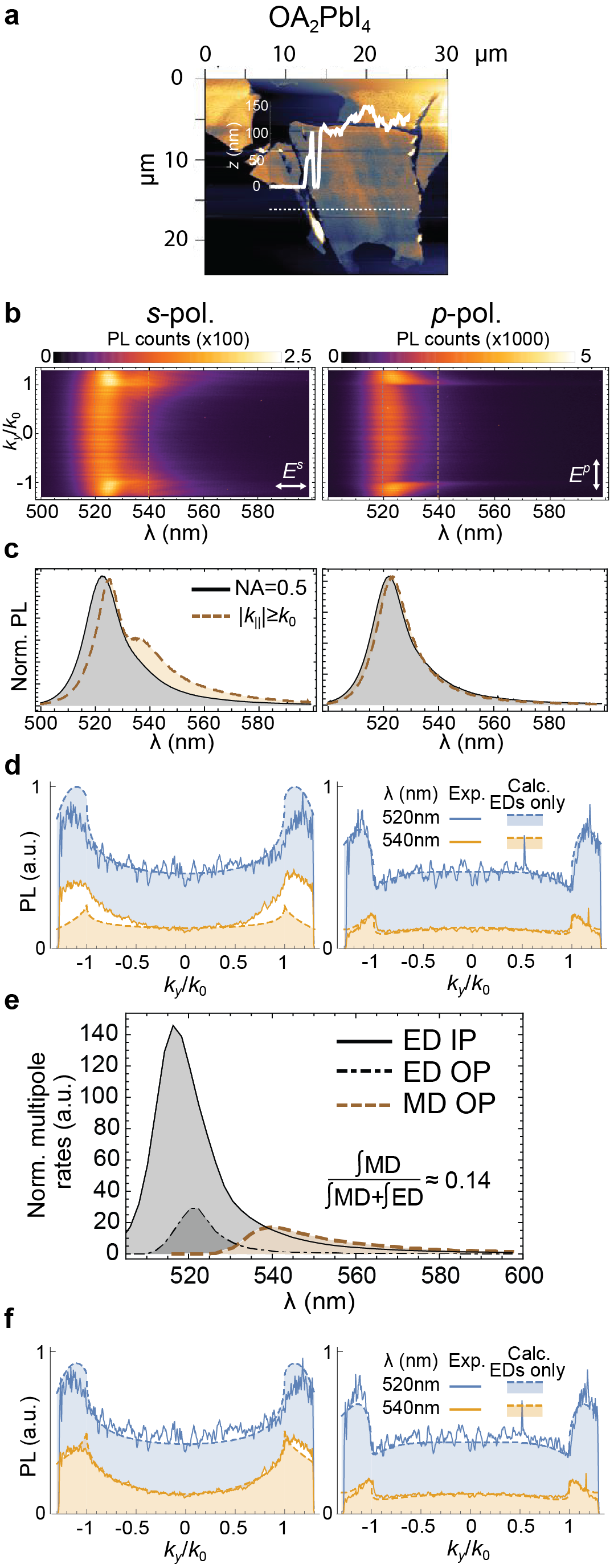}
		\caption{
		\textbf{\emph{Summarized analysis of exfoliated single crystals of OA$_2$PbI$_4$.}}
		\footnotesize
		(All subpanel descriptions identical to those of Supplementary Fig. \ref{fig:flake_BA}.) Optical constants were were derived from a layered effective medium model applied to thin films of \BAPbI.
		}
	\end{minipage}
	\label{fig:flake_OA}
\end{figure}

\newpage
\begin{figure}[H]{}
	\centering
	\includegraphics[width=0.45\linewidth]{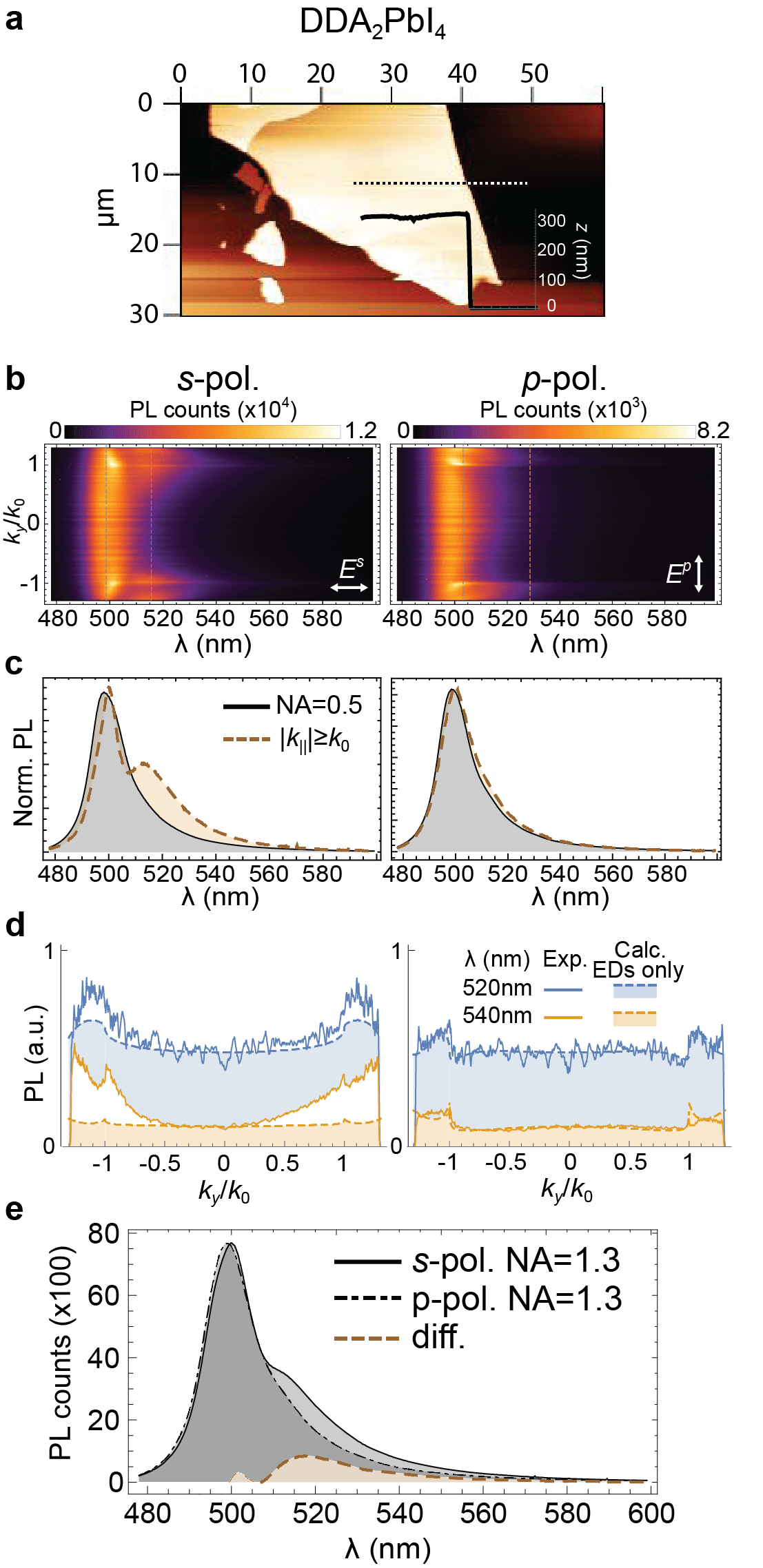}
	\caption{
	\textbf{\emph{Summarized analysis of exfoliated single crystals of DDA$_2$PbI$_4$.}}
		\footnotesize
		(\textbf{a}-\textbf{d}) (Subpanel descriptions identical to those of Supplementary Fig. \ref{fig:flake_BA}.)
		(\textbf{e}) Momentum-integrated (NA=1.3) $s$- (black solid) and $p$-polarized (black dashed) PL spectra. The $p$-polarized spectrum has been normalized to give a peak height at 520 nm equal to that of the $s$-polarized spectrum. The difference spectrum is represented by the brown dashed curve. Note that the primary exciton PL peak is at 500 nm. Rigorous multipolar decompositions were not performed on this sample since the optical constants are not well known.
		}
	\label{fig:flake_DDA}
\end{figure}

\newpage
\section{Photo-degradation effects upon the multipolar emission energy}
\begin{figure}[H]{}
\centering
	\includegraphics[width=0.8\linewidth]{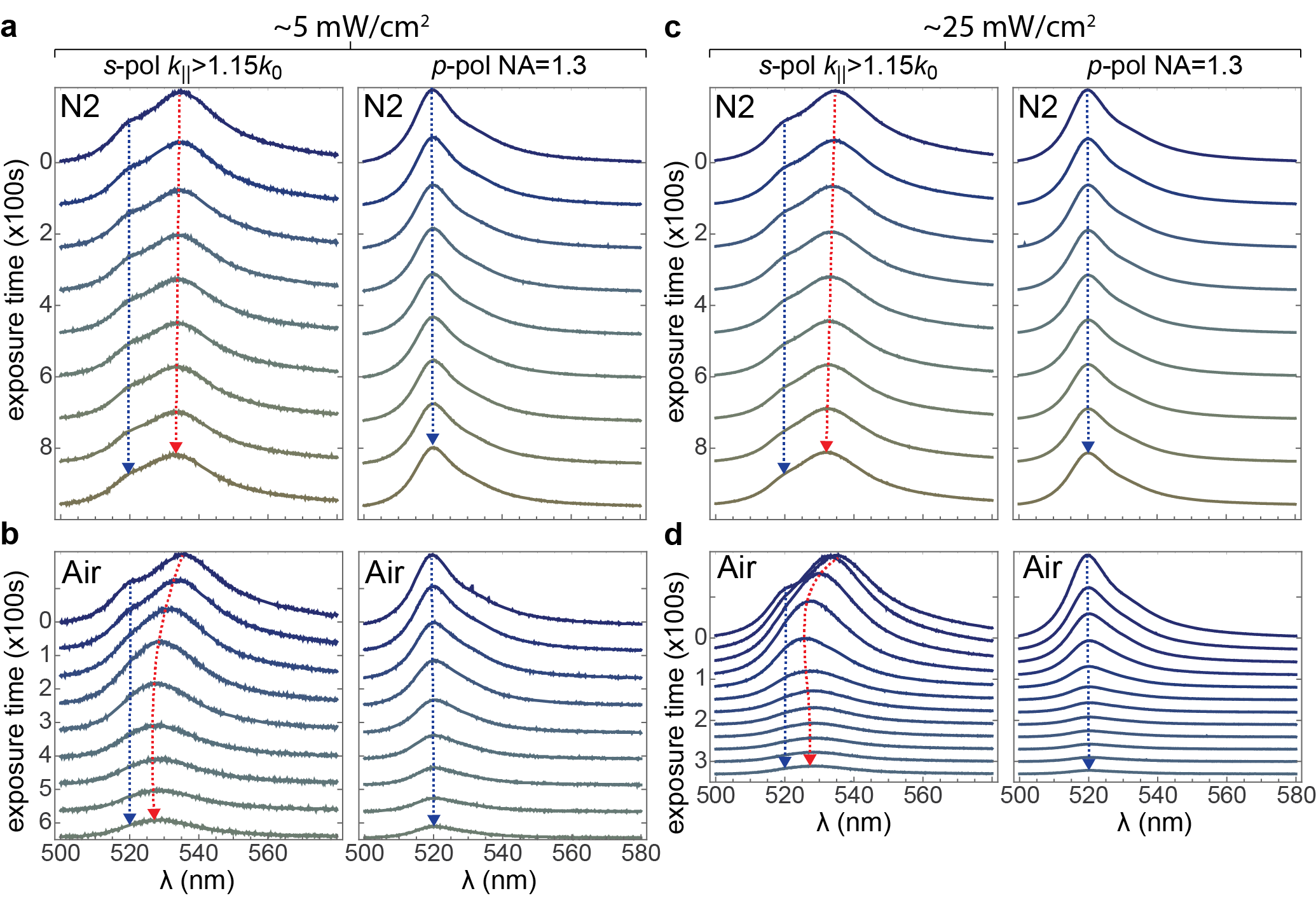}
	\caption[]{
	\textbf{\emph{Illumination intensity- and environment-dependent evolution of EQ emission in thin films of BA$_2$PbI$_4$}}  
	Temporal evolution of polarized emission spectra with varying illumination intensity in (\textbf{a},\textbf{c}) inert (nitrogen, `N2') and (\textbf{b},\textbf{d}) ambient (`air') environments.  
	(\textbf{a},\textbf{b}) $s$-Polarized (left plots) and $p$-polarized (right plots) PL spectra in inert (nitrogen, `N2') and ambient (`air') environments under 5 mW/cm$^2$ incident power.  $s$-Polarized spectra were collected at $|k_{||}|$$>$1.15$k_0$.  $p$-Polarized spectra are integrated over NA=1.3.  Red (blue) dashed arrows trace the spectral location of the EQ (ED) emission feature through time.
	(\textbf{c},\textbf{d}) Same as (a,b), but for an incident power of 25 mW/cm$^2$.
	}
	\label{fig:environment}
\end{figure}

Figure \ref{fig:environment} shows the results of photo-induced degradation studies as described in the Methods section of the manuscript. Here we use a cascaded representation in which the vertical axis represents total illumination exposure times (in units of 100~s), beginning from $t$=0 (top; illumination `turned on') and evolving `downward'.  Polarization- and momentum-dependent PL spectra were recorded in 10 s increments, beginning at $t$=0 and continuing until net PL became negligible.  A uniformly-spaced subset of PL spectra are plotted with a zero-point at a height corresponding to the time the exposure was started.  (A subtle color gradient is implemented to aid interpretation of the time evolution.)  PL spectra are shown for both $s$-polarization at $|k_{||}|>1.15k_0$ (left plots in each subfigure) and $p$-polarization integrated over NA=1.3 (right plots in each subfigure). In an N$_2$ environment (Figs. \ref{fig:environment}a,c), thin films of BA$_2$PbI$_4$ exhibit high photostability; the PL spectra remained virtually constant for up to several hours of exposure even under significant 25 mW/cm$^2$ illumination intensities. While the net PL intensity is virtually undiminished with time (vertical axis), we observe a slight blue-shifting of the EQ emission feature with exposure time, as observed in high-$k_{||}$ $s$-polarized PL spectra (left plots; red dashed lines).  The rate of blue-shift increases with illumination intensity. In contrast, the peak wavelength of the ED emission feature remains unchanged (blue dashed lines), as observed in $p$-polarized spectra (right plots). In air (Figs. \ref{fig:environment}b,d), rapidly diminishing PL intensity reflects efficient photo-induced degradation of the thin films, likely caused by the presence of oxygen of water in ambient conditions \cite{aristidou_role_2015, marinova_hindered_2017}. From this, we conclude that the observed variation of the multipolar emission energy in similarly prepared samples arises from variations in processing conditions as well as the age of the sample. Note that the degradation observed here is not reversible.
	
\newpage
\section{Radiation patterns from oriented multipoles}\label{Sect:multipoles}
\begin{figure}[H]
\begin{center}
  \includegraphics[width=\linewidth]{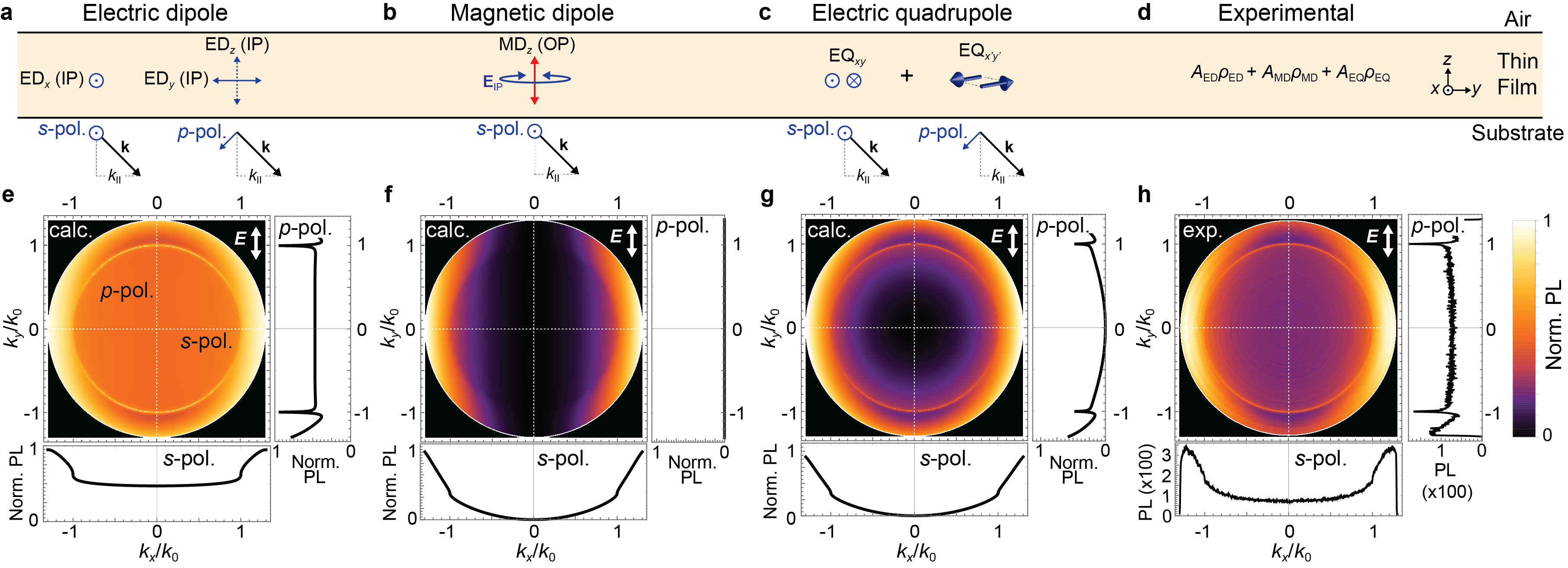}
  \caption{
  \textbf{\emph{BFP radiation patterns from oriented multipoles.}} 
	\footnotesize(\textbf{a-c}) Schematic illustrations of the oriented multipolar candidates consistent with our observations: (\textbf{a}) in-plane (IP) and out-of-plane (OP) EDs (OP:IP ED ratio of 0.07); (\textbf{b}) an OP MD; (\textbf{c}) IP transverse EQs. 
	(\textbf{d}) Experimental radiation patterns are linear combinations of terms considered in subpanels a-c.
	(\textbf{e-g}) Calculated $y$-polarized momentum-resolved luminescence patterns from (\textbf{e}) EDs, (\textbf{f}) an OP MD, and (\textbf{g}) IP EQs.
	(\textbf{h}) Experimental BFP image.
	$s$-Polarized and $p$-polarized traces are shown below and to the right of each 2D image, respectively. All calculations were performed for a 61 nm film using optical constants from Supplementary Fig. \ref{Fig:Constants}.
	}
  \label{Fig:BFP_1}
  \end{center}
\end{figure}

\begin{figure}[H]
\begin{center}
  \includegraphics[width=0.65\linewidth]{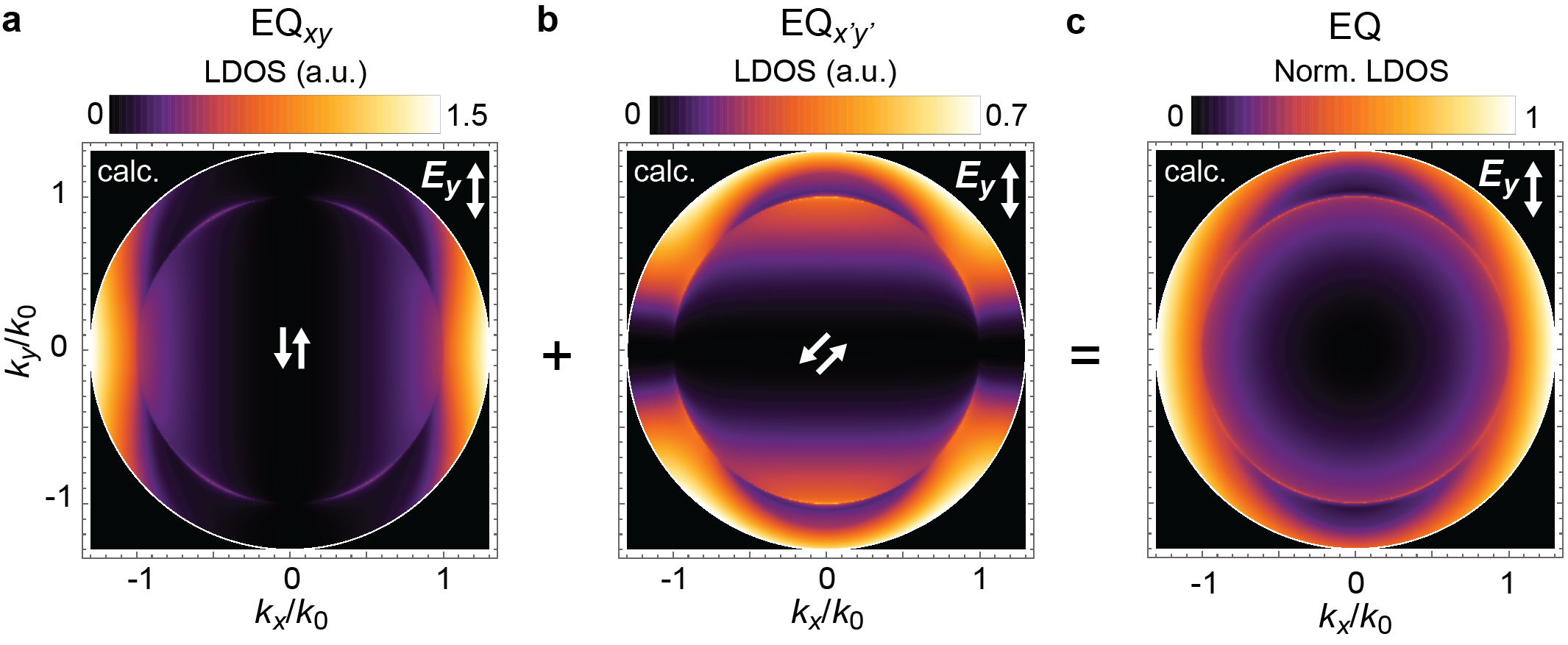}
  \caption{\textbf{\emph{BFP radiation patterns from in-plane EQs.}} 
	\footnotesize
	(\textbf{a}-\textbf{b}) Calculated $y$-polarized (normalized) LDOS for (\textbf{a}) ``EQ$_{xy}$" and (\textbf{b}) ``EQ$_{x'y'}$". 
	(\textbf{c}) The sum of the two distinct radiation patterns presented in \textbf{a}-\textbf{b} represents IP EQ radiation from a rotationally invariant system.}
  \label{Fig:BFP_2}
  \end{center}
\end{figure}

Here we provide a detailed description of the radiation characteristics of the oriented EDs, MDs, and EQs considered in the manuscript. The multipolar species are described schematically in Fig. \ref{Fig:BFP_1}a-c with accompanying LDOS calculations shown below (Figs. \ref{Fig:BFP_1}e-g). The experimental 540 nm BFP image is shown in Fig. \ref{Fig:BFP_1}h. In the lower subpanels, $s$- and $p$-polarized $\kp$-space linecuts are shown below and to the right of each LDOS image, respectively. For spin-cast films which exhibit rotational symmetry about the $z$-axis, all IP terms must be included with equal weight. IP EDs (``ED$_x$") contribute to $s$-polarized emission, while both IP (``ED$_y$") and OP (``ED$_z$") EDs contribute to the $p$-polarized emission. In contrast, OP MDs (``MD$_z$") are associated with a circulating IP electric field (blue circular arrow) and thus only produce $s$-polarized emission. Two distinct IP EQ terms (``EQ$_{xy}$" and ``EQ$_{x'y'}$") must be considered. The individual EQ contributions are illustrated in Fig. \ref{Fig:BFP_2}; EQ$_{xy}$ contributes symmetrically to PL along $k_y$=0 and contributes nothing to $k_x$=0. (The $k_y$=0 line cut is functionally identical to that of an out-of-plane MD, as understood by the electric field symmetries associated with the two multipolar species.) The same in-plane EQ rotated by 45$^\circ$, EQ$_{x'y'}$, is distinct, exhibiting a ``four-fold" symmetry in the back focal plane (four equivalent emission maxima, one in each quadrant). Both EQ basis elements must be included in rotationally invariant systems. The various EQ configurations are schematically represented in the center of each BFP image of Figs. \ref{Fig:BFP_2}a-b. The LDOS is presented in arbitrary units, but the relative scaling is retained between Figs. \ref{Fig:BFP_2}a-b.

\newpage
\section{TRPL and pump power-dependence of multipolar decomposition in thin films}\label{Sect:Pump}
\begin{figure}[H]
\begin{center}
  \includegraphics[width=0.5\linewidth]{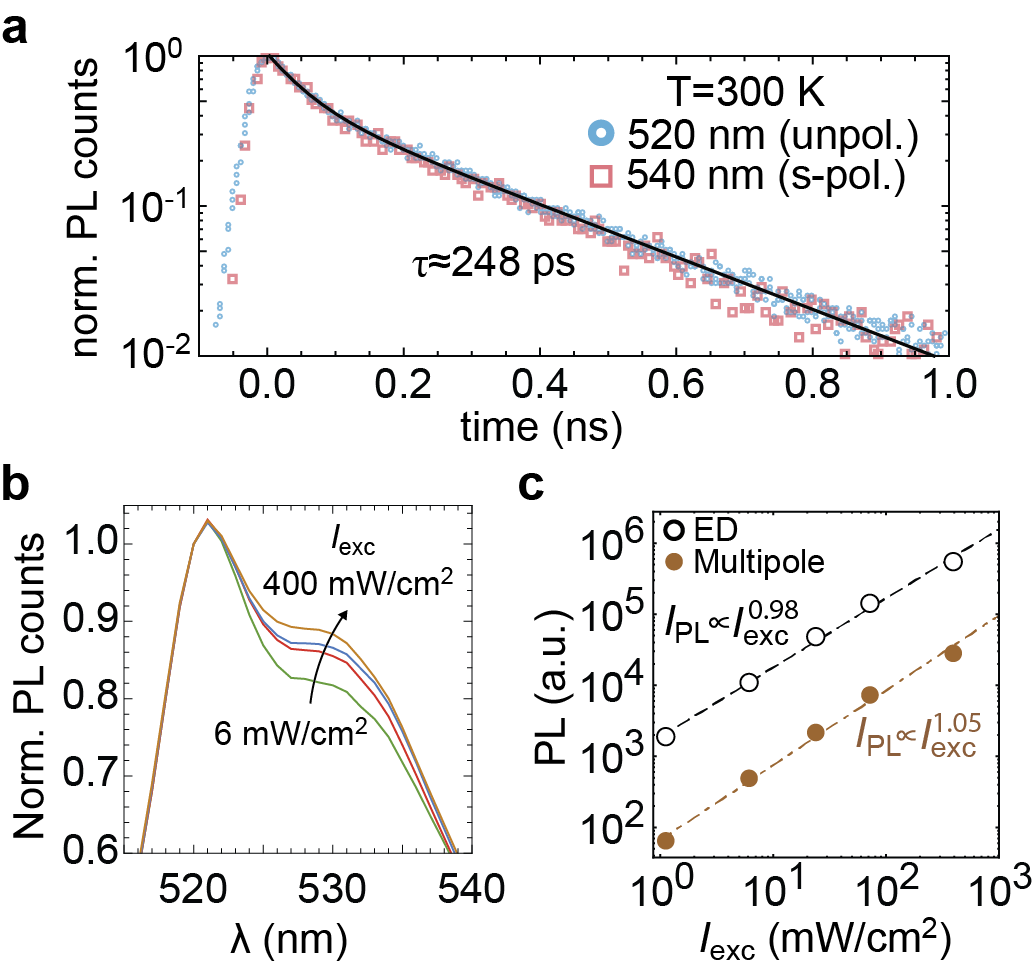}
  \caption[]{\textbf{\emph{TRPL and pump power-dependence of multipolar decomposition in spin-cast thin films of \BAPbI.}} 
	(\textbf{a}) PL decay traces collected at 520 nm (unpolarized) and 540 nm ($s$-polarized), showing sub-ns lifetimes comparable to those observed in drop-cast samples. No discernible difference is observed between the two decay traces.
    (\textbf{b}) High-momentum ($\kp$$>$$k_0$) PL spectra for various excitation intensities ($I_\text{exc}$$\approx$6-400 mW/cm$^2$). 
    (\textbf{c}) Power-law fits ($I_\text{PL}$$\propto$$I_\text{exc}^\alpha$) of the integrated multipolar strengths, (black) $I_\text{PL}$=$\int A_\text{ED}(\lambda) d\lambda$ and (brown) $I_\text{PL}$=$\int A_\text{MD}(\lambda) d\lambda)$, as a function of excitation intensity, $I_\text{exc}$. Note that the sample used for subpanels (b)-(c) shows the strong polarization-dependent shoulder at $\approx$530 nm. PL spectra (b) show that the 530 nm shoulder grows in relative strength over the 520 nm feature at high momenta. However, the momentum-integrated strength of the ED contribution (black; $\alpha_\text{ED}$=0.98$\pm$0.04) and the multipolar contribution (brown; $\alpha_\text{MD}$=1.05$\pm$0.05) are both consistent with a linear pump-power dependence ($I_\text{PL}$$\propto$$I_\text{exc}$), thereby supporting interpretation as excitonic (vs. bi-excitonic) emission. The observed relative increase in shoulder emission strength may thus be a consequence of a redistribution of emission power into higher momenta due to, e.g., pump-dependent changes in the film refractive index.}
  \label{Fig:TRPL_Pump}
  \end{center}
\end{figure}

Figure \ref{Fig:TRPL_Pump} shows TRPL and pump power-dependence studies performed on spin-cast thin films of \BAPbI, analogous to Fig. 4 of the manuscript. Thin film measurements presented here were performed at 300 K. To maximize the signal from the 540 nm emission feature, PL was collected at oblique exitance ($\sim$50$^\circ$) with a linear polarizer corresponding to an $s$-polarized collection configuration. 

\newpage
\section{Numerical band-structure calculations and charge densities}\label{Sect:DFT}
\begin{figure}[H]
\begin{center}
  \includegraphics[width=0.98\linewidth]{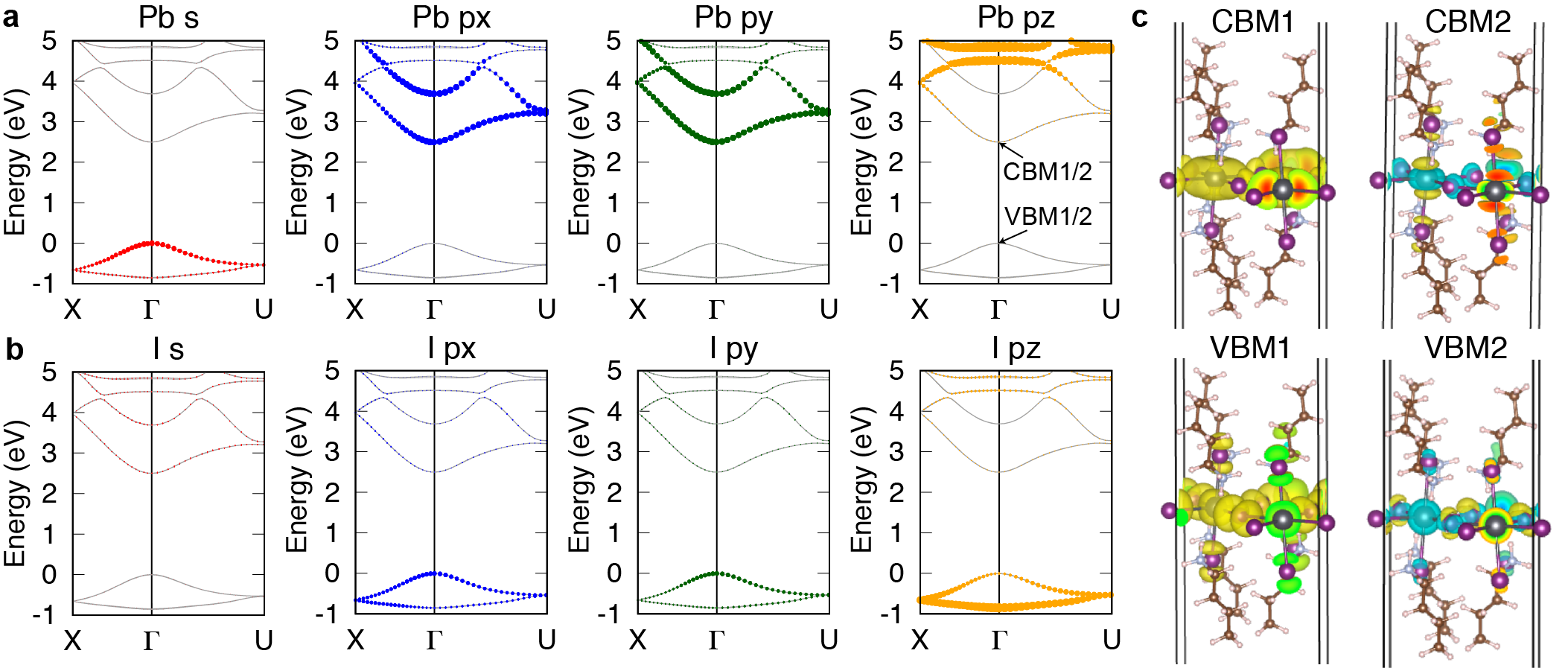}
  \caption[]{\textbf{\emph{Calculated band structure, character, and charge densities.}} 
	(\textbf{a}) Band structure including wave function projections onto Pb orbitals with symmetries corresponding to $s$, $p_x$, $p_y$, and $p_z$. 
	(\textbf{b}) Band structure including wave function projections onto I orbitals with symmetries corresponding to $s$, $p_x$, $p_y$, and $p_z$.
	Both the valence band maximum (VBM) and conduction band minimum (CBM) are doubly degenerate within the current treatment including spin-orbit coupling.
	(\textbf{c}) 3D electronic charge densities associated with the two degenerate (top) CBM states and (bottom) VBM states at $\Gamma$.}
  \label{Fig:DFT}
  \end{center}
\end{figure}

\newpage
\section{Matrix elements of multipolar transitions}\label{Sect:MatrixElements}
Here we discuss the multipolar optical transition selection rules by describing the initial (CBM) and final (VBM) states in terms of Bloch functions (at $\mathbf{k}$=0) with $p$- and $s$-like symmetries, respectively. We refer to this as the ``conventional treatment", in analogy with the formalism developed for III-V semiconductors,\cite{bastard_wave_1988} as used in several recent reports.\cite{even_electronic_2012,becker_bright_2018} We show that within this treatment MD and EQ transitions are strictly forbidden. \\

In the presence of spin-orbit coupling (SOC), neither orbital ($L$) nor spin ($S$) angular momentum are good quantum numbers, and one instead takes eigenstates of \emph{total} angular momentum (squared), $\hat{J}^2 = \hat{\mathbf{J}}^2 = (\hat{\mathbf{L}} + \hat{\mathbf{S}})^2$, and its projection, $\hat{J}_z$, along a particular quantization axis: $\ket{J,J_z}$ \cite{bastard_wave_1988,even_electronic_2012,becker_bright_2018}. The quantization axis, $\hat{z}$, is conveniently taken to be the dimension in which carriers are most strongly confined (e.g., the $c$-axis in 2D HOIPs or the out-of-plane direction for nanoplatelets of 3D HOIPs). In HOIPs, the CBM is associated with the spin-split band with $J$=1/2.  In previous studies, the VBM is treated with an overall $s$ symmetry, and is thus represented by the Bloch function $\ket{S}$ ($L$=0, and thus $J$=1/2) \cite{tanaka_electronic_2005,even_electronic_2012,becker_bright_2018}. The orthonormal CBM and VBM states within this description are:
\ba{
\ket{\psi_{c,+}} &= \left|\half, +\half \right\rangle_c  = 
\frac1{\sqrt2} (\ket{X}_c + i\ket{Y}_c)\ket{\down} \label{eqn:CBM1}  \\ 
\ket{\psi_{c,-}} &= \left|\half, -\half \right\rangle_c = 
- \frac1{\sqrt2} (\ket{X}_c - i\ket{Y}_c)\ket{\up} \label{eqn:CBM2} \\ 
\ket{\psi_{v,+}} &= \left|\half, +\half \right\rangle_v = \ket{S}_v \ket{\up} \label{eqn:VBM1}  \\ 
\ket{\psi_{v,-}} &= \left|\half, +\half \right\rangle_v = \ket{S}_v \ket{\down} \label{eqn:VBM2} 
}

in which $v$ ($c$) denotes valence (conduction) band states, and $\ket{\up}$ ($\ket{\down}$) is the eigenstate of $\hat{\mathbf{S}}$ with eigenvalue +1/2 (-1/2). $\ket{X}$ and $\ket{Y}$ correspond to real $p$-like orbitals (analogous to atomic $L$=1 orbitals) with symmetry axes along the Cartesian $x$ and $y$ directions, respectively.  We have here assumed that the $\ket{Z}$ contribution at the CBM is negligible, in accord with the 2D nature of the orbitals\cite{even_electronic_2012} and DFT calculations presented in the manuscript. By symmetry considerations, it is immediately apparent that the ED matrix elements, $M_{ED} \propto \bra{1/2,\pm1/2} \hat{x}_i \ket{1/2,\pm1/2}$ ($i$=1,2,3), are \emph{in general} non-zero since the individual terms $\bra{S}_v \hat{x} \ket{X}_c$=$\bra{S}_v \hat{y} \ket{Y}_c$$\neq$0, thus driving in-plane electric dipole band-to-band transitions in HOIPs. Accounting for electron-hole exchange (i.e., excitons), specific singlet/triplet combinations may become ED-allowed or ED-forbidden, as determined by considering the proper product combinations of states as presented in Eqns. \ref{eqn:CBM1}-\ref{eqn:VBM2}; e.g., $\ket{\psi_{c,i}}\ket{\psi_{c,j}}$ with $i,j$=+,$-$.\cite{tanaka_electronic_2005,becker_bright_2018} (Equivalently, there are four distinct matrix elements to consider for the states in Eqns. \ref{eqn:CBM1}-\ref{eqn:VBM2}, some of which are zero.) \\

\subsection{Magnetic dipole transitions}
The MD vector operator is given by $\hat{\mathbf{m}}$=$(e/2mc)(\hat{\mathbf{L}} + 2 \hat{\mathbf{S}})$ where $e$ and $m$ are the charge and mass of the particle in question, respectively.\cite{bunker_molecular_2006} Our experimental results motivate us, in particular, to inspect $\hat{m}_z$$\propto$$\hat{L}_z$+$2 \hat{S}_z$, and thus we analyze matrix elements of the form $\bra{\psi_{v,i}} \hat{m}_z \ket{\psi_{c,j}}$$\propto$$\bra{\psi_{v,i}} (\hat{L}_z + 2 \hat{S}_z) \ket{\psi_{c,j}}$ with $i,j$=+,$-$. Notice that $\ket{\psi_{v,i}}$ and $\ket{\psi_{c,j}}$, as written in Eqns. \ref{eqn:CBM1}-\ref{eqn:VBM2}, are superpositions of $\hat{L}_z$ eigenstates, all terms in which exist within a subspace associated with a single specific value of $L$; i.e., $L$=0 for VBM states and $L$=1 for CBM states. (More generally, all terms exist within a subspace associated with a single specific \emph{parity} of $L$; e.g., $L$=0,2,4,... for VBM states. These are, however, small contributions and rigorously do not affect the conclusions of our following arguments.) Therefore, all terms $\bra{\psi_{v,i}} \hat{m}_z \ket{\psi_{c,j}}$ are immediately seen to be zero. Specifically, restraining our attention to $L$=0 and $L$=1 leading terms for the VBM and CBM, respectively, Eqns. \ref{eqn:CBM1}-\ref{eqn:CBM2} can be rewritten in terms of spherical harmonics ($\ket{l,m_l}$):
\ba{
\ket{\psi_{c,+}} &\approx \ket{L=1,m_l=+1}_c \ket{\down} \label{eqn:CBM1_2} \\
\ket{\psi_{c,-}} &\approx - \ket{L=1,m_l=-1}_c \ket{\up} \label{eqn:CBM2_2}
}

and $\ket{S}_v$$\approx$$\ket{L=0,m_l=0}_v$. (The symbol $\approx$ is used to indicate that we have chosen to ignore higher values of $L$ with equivalent parity, according to the statement in the previous paragraph.) With these we analyze the terms in the matrix elements of interest:
\ba{
\bra{\psi_{v,i}} (\hat{L}_z + 2 \hat{S}_z) \ket{\psi_{c,j}} = 
\bra{\psi_{v,i}} \hat{L}_z \ket{\psi_{c,j}} +2  \bra{\psi_{v,i}} \hat{S}_z \ket{\psi_{c,j}}
= j \braket{\psi_{v,i}}{\psi_{c,j}} + (-j) \braket{\psi_{v,i}}{\psi_{c,j}} = 0
}

where $j$=$+1$,$-1$ and we have utilized the orthogonality of the terms with differing $L$. Since $\hat{L}_z \ket{S}$=0, we see that an odd-parity term in the VBM is sufficient to generate a non-zero $\hat{m}_z$ matrix element. In particular, since $\hat{L}_z \ket{S}$=0, we conclude that odd-parity terms must be included in the VBM.  The preceding arguments remains true even when accounting for excitonic correlations of the electron and hole states, specifically from the 1s exciton state, as this can only introduce a spherically (or, in 2D, circularly) symmetric envelope function to the analysis. That is, the multipolar character of the transition is determined from the underlying electron and hole Bloch states comprising the exciton \cite{tanaka_electronic_2005}. (Note that this selection rule may be relaxed for $p$-like exciton states, but these states lie at energies $\approx$300 meV \emph{higher} than than the 1s exciton state \cite{tanaka_electronic_2005}.)

\newpage

\bibliographystyle{bibstyle_NatMat}
\bibliography{supp_bib}